\PassOptionsToPackage{table}{xcolor}
\documentclass[a4paper,twocolumn,11pt]{quantumarticle}
\pdfoutput=1
\usepackage[utf8]{inputenc}
\usepackage[english]{babel}
\usepackage[T1]{fontenc}
\usepackage{amsmath}
\usepackage[numbers]{natbib}

\usepackage{url}            
\usepackage{booktabs}       
\usepackage{amsfonts}       
\usepackage{nicefrac}       
\usepackage{microtype}      
\usepackage{xcolor}         
\usepackage{acronym}
\usepackage{braket}
\usepackage{algorithm}
\usepackage[noend]{algpseudocode}
\usepackage{multirow}
\usepackage{mathtools}
\usepackage{bm}
\usepackage{tikz}
\usepackage{tabularx}      
\usepackage{subcaption}
\usepackage{hyperref}
\usepackage{cleveref}

\DeclareMathOperator*{\argmax}{argmax}
\newcommand{\valcell}[2]{\cellcolor{green!#1!red}#2}

\begin{document}

\title{A Multiclass Quantum Aligned Centroid Kernel}

\author{Kilian Tscharke}
\affiliation{Fraunhofer Institute for Applied and Integrated Security (AISEC), 85748 Garching near Munich, Germany}
\orcid{0009-0006-7423-2498}
\email{firstname.lastname@aisec.fraunhofer.de}

\author{Pascal Debus}
\affiliation{Fraunhofer Institute for Applied and Integrated Security (AISEC), 85748 Garching near Munich, Germany}
\orcid{0009-0007-2437-2764}
\maketitle

\begin{abstract}
 Kernel methods are powerful tools in machine learning but commonly used full-Gram kernels face three key limitations: (1) quadratic scaling with training set size; (2) the use of fixed, non-trainable kernels; and (3) the absence of an intrinsic formulation for multiclass classification. 
We present McQuack, a trainable quantum kernel method for multiclass problems that achieves linear scaling in the number of training samples. 
This is accomplished by replacing the full training-set Gram matrix with a trainable sample-to-(class-centroid) fidelity matrix.
We evaluate the model in simulation and on 124 qubits of two IBM devices, across more than 150 datasets. 
In simulation, McQuack outperforms existing "pure" quantum baselines, while results from hardware inference -- obtained without training -- achieve performance similar to an RBF kernel. 
Finally, we study the trainability of the model and observe no evidence of barren plateaus in our experiments with up to 13 qubits, and highlight the importance of parameter initialization for successful optimization.
\end{abstract}

\section{Introduction}

\acp{KM} are effective in high-dimensional spaces but common full-Gram kernels suffer from three key limitations:
First, the construction of the kernel matrix scales as $\mathcal{O}(n_\text{train}^2)$ for a training set of size $n_\text{train}$.
Second, often used kernels such as the \ac{RBF} are fixed and cannot adapt to the data.
Third, \acp{KM} are typically designed for binary classification, while multiclass extensions (e.g., one-vs-one \acp{SVC}\footnote{\url{https://scikit-learn.org/stable/modules/generated/sklearn.svm.SVC.html}}) can incur an $\mathcal{O}(M^2)$ training cost for $M$ classes.

\acp{QKM} utilize quantum feature spaces to address these limitations.
They can employ trainable embedding kernels optimized via \ac{KTA} \cite{quantum_kernel_alignment_2022}, and have been shown to offer advantages in specific constructed learning tasks \cite{quantum_kernel_advantage_discrete_log_2021, Huang_2021}.

Despite promising theoretical results, evaluating the practical potential of \ac{QML} remains challenging due to hardware limitations, experimental bias (i.e. the desire of researchers to publish methods that perform better than existing ones), and the need for statistically robust benchmarks \cite{schuld_qml_benchmarks2024}.
Thus, extensive and critical benchmarking of new \ac{QML} approaches is essential. 
Another major challenge when scaling up \ac{QML} models is trainability. Barren plateaus -- regions of the optimization landscape in which gradient magnitudes or variances decrease exponentially with the number of qubits -- severely hinder gradient-based optimization \cite{Barren_Plateaus_McClean_2018}. Unfortunately, hardware-efficient ansätze commonly used in the NISQ-era, such as data re-uploading circuits, are also susceptible to this effect \cite{larocca2024reviewbarrenplateausvariational}. 
However, recent evidence suggests that certain initialization strategies (e.g. warm-starting) can mitigate barren plateaus. Therefore, we investigate different initialization strategies to improve the trainability of our models.


The remainder of this paper is structured as follows:
The next subsection \ref{Related Work} gives an overview of the related work in the fields of \acp{QKM}, benchmarking \ac{QML}, and \ac{KTA}.
The final part of the introduction highlights our contributions (subsection \ref{Contributions}).
The following Background (section \ref{Background}) introduces the fundamentals of multiclass classification and trainable quantum kernels.
Afterwards, section \ref{sec:McQuack} explains the theoretical foundation of our model.
In the Results and Discussion (section \ref{Results and Discussion}) we show the results of our experiments and analyze them.
Finally, Conclusion and Outlook (section \ref{Conclusion and Outlook}) highlights the key results of this work and gives future research directions.
The detailed description of the model and the datasets, as well as the experimental details, and an introduction to \acp{QKM} can be found in the appendices.

\subsection{Related Work} \label{Related Work}

Tscharke et al. \cite{Quack} introduced Quack, a quantum kernel-based algorithm for binary classification with linear runtime complexity in the number of training samples. However, their work left several gaps open, most notably the lack of experimental validation on quantum hardware and the generalization to multiclass settings. Both of these limitations are addressed in the present work. \\
Several works have investigated \acp{QKM} optimized via \ac{KTA} for binary classification \cite{Glick_2024,Sahin_2024,quantum_kernel_alignment_2022,QKA_w_stoch_grad_desc_Gentinett_2023,kruse_kernel_alignment,Pellow_Jarman_2024_kernel_alignment}. While these approaches demonstrate improved performance through trainable feature maps, extensions to intrinsically multiclass formulations remain limited. Existing multiclass approaches typically rely on one-vs-all strategies  \cite{QKM_for_multiclass_neuron_classification,bitflip_tolerance}, which result in additional computational overhead. 
In contrast, \textsc{McQuack} provides a native multiclass formulation with linear scaling in the number of training samples and classes.
Agliardi et al. \cite{bitflip_tolerance} introduced \emph{bit flip tolerance}, a pragmatic noise-mitigation strategy designed for fidelity-based kernels that is particularly well suited for \textsc{McQuack}. \\
A significant contribution to the benchmarking of \ac{QML} models was made by Bowles et al. \cite{schuld_qml_benchmarks2024}. Their work highlights key challenges in the fair evaluation of these algorithms, and provides a comprehensive benchmarking suite for assessing new \ac{QML} models. We use their framework to determine the potential of our model.

\subsection{Contributions} \label{Contributions}
Our work aims to evaluate the potential of our Multiclass Quantum Aligned Centroid Kernel (\textsc{McQuack}) algorithm for (multiclass) classification tasks. Our contributions are as follows:

\begin{enumerate}
    \item We introduce \textsc{McQuack}, a multiclass \acl{QKM} that optimizes a trainable kernel via \acl{KTA} and achieves linear runtime complexity with respect to the training set size and number of classes.
    \item We evaluate the model without prior training on 124-qubit subgraphs of two 156-qubit IBM devices and demonstrate performance comparable to an \ac{RBF} kernel despite hardware noise.
    \item We conduct extensive benchmarking against classical and quantum baselines on the \ac{QML} benchmark suite \cite{schuld_qml_benchmarks2024} comprising 133 synthetic binary datasets and additionally, evaluate the model on 19 real-world datasets with up to ten classes, showing that \textsc{McQuack} outperforms existing "pure" quantum baselines\footnote{By "pure" quantum models we refer to models without trainable classical components, e.g. without neural networks for pre-/postprocessing.}.
    \item We analyze the trainability of the model and observe no evidence of barren plateaus for up to 13 qubits, and highlight the importance of the parameter initialization.
\end{enumerate}

\section{Background} \label{Background}
This section introduces the fundamentals of multiclass classification, trainable kernels, and \ac{KTA}. 
This provides the necessary background for understanding the \textsc{McQuack} algorithm described in \cref{sec:McQuack}.
For the reader unfamiliar with \acp{QKM}, an introduction can be found in \cref{sec:appD:quantum_kernels}.

\subsection{Multiclass Classification}
Let $\mathcal{X}\subset\mathbb{R}^d$ be the data space and 
$\boldsymbol{x_i} = (x^{(1)},\ldots,x^{(d)}) \in \mathcal{X}$ the feature vector of sample $i$. 
Let $\mathcal{Y} = \{0,1,\ldots,M-1\}$ denote the target space for $M$ classes and $y_i \in \mathcal{Y}$ the label of a single sample. 
Let $\Theta$ further denote the space of model parameters and $\boldsymbol{\theta} \in \Theta$ the model parameters.
The task of \textit{supervised multiclass} classification is to train a parameterized model $f: \mathcal{X} \times \Theta\to\mathcal{Y}$ such that it approximates a mapping between inputs $\boldsymbol{x_i}$ and labels $y_i$ based on the learned parameters $\boldsymbol{\theta}$, as described in \eqref{eq_output_ad}.
During training,$\boldsymbol{\theta}$ is optimized  according to \eqref{eq_loss_ad} such that a loss $L$  quantifying the difference between the predicted output $\hat{y}$ and the ground truth $y$ is minimized.
\begin{align}
    \hat{y} = f (\boldsymbol{x}; \boldsymbol{\theta}) \label{eq_output_ad} \\
    \min_\theta L(y, \hat{y}) \label{eq_loss_ad}
\end{align}

\subsection{Trainable Quantum Kernels} \label{Trainable Quantum Kernels}
A kernel is a real- or complex-valued positive-definite function of two data points, i.e. $\kappa : \mathcal{X} \times \mathcal{X} \rightarrow \mathbb{K}$, where $\mathbb{K} \in \{\mathbb{R}, \mathbb{C}\}$.
This definition can be extended to the quantum case, where the quantum kernel $k$ between two data-encoding pure quantum states $\psi (\boldsymbol{x_i}) = \psi_{i}$ and $ \psi (\boldsymbol{x_j}) = \psi_{j}$ is calculated from the fidelity $F$ between these states
\begin{align}
     k(\boldsymbol{x_i}, \boldsymbol{x_j}) &= F(\psi_{i}, \psi_{j}) = \vert \braket{\psi_{j}|\psi_{i}} \vert ^2 \notag \\
     &= \vert \bra{0^{\otimes n}} U^\dagger(\boldsymbol{x_j}) U(\boldsymbol{x_i}) \ket{0^{\otimes n}} \vert ^2 \label{eq:quantum_kernel_hardware1}
\end{align}
with data encoding unitaries $U(\boldsymbol{\cdot})$ and its conjugate transpose $U^\dag(\boldsymbol{\cdot})$.
Positive definiteness of this kernel has been proven in \cite{schuld2021supervisedquantummachinelearning}.
\Cref{eq:quantum_kernel_hardware1} shows how the input data are encoded into the kernel, however, quantum kernels may also include trainable parameters that can influence their performance on a given dataset. 
A common approach is to encode the input $\boldsymbol{x'}\in\mathbb{R}^3$ in $U$ by using a rotation gate  
$R\!\left(\boldsymbol{w}_v \circ \boldsymbol{x'} + \boldsymbol{b}_v\right),$
where the weights $\boldsymbol{w}_v\in\mathbb{R}^3$ and bias $\boldsymbol{b}_v\in\mathbb{R}^3$ are optimized during training. For general $\boldsymbol{x}\in\mathbb{R}^d$, these variational embedding gates can be stacked arbitrarily by adding either more qubits or more layers (hence the index $v$). In this way, individual input features can be re-encoded multiple times, a technique commonly referred to as \textit{data re-uploading} \cite{reupload_encoding_2020}.

This encoding structure also offers a theoretical benefit: a quantum model with a unitary consisting of a data encoding and a trainable block can be expressed as a partial Fourier series \cite{effect_of_data_encoding_Schuld_2021}. With $r$ repetitions of the unitary, the model corresponds to a truncated Fourier series of degree $r$, which means that the number of accessible Fourier frequencies grows linearly with the number of data re-uploading repetitions.

\subsection{Quantum Kernel Target Alignment for Non-Gram Matrices} \label{Quantum Kernel Alignment}
\ac{KTA} is a similarity measure that quantifies how well the estimated matrix $K$ captures the distribution of the data, in relation to the ideal matrix $K^*$.
Originally developed for classical kernels \cite{kernel_alignment_2001}, it has since been extended to quantum kernels \cite{quantum_kernel_alignment_2022}. Higher \ac{KTA} correlates with improved performance of kernel-based classifiers.

In the multiclass setting relevant for \textsc{McQuack}, the kernel is a sample-to-(class-centroid) matrix rather than a Gram matrix. For $n_{\text{samples}}$ samples, the ideal matrix $K^* \in \{-1,1\}^{n_{\text{samples}} \times M}$ encodes the relationship between sample $i \in \{0, \dots, n_{\text{samples}}-1 \}$ and class $m\in \{0, \dots, M-1 \}$ as
\begin{align}
    K_{i,m}^* = 
    \begin{cases} 
        1 & \text{if } y_i = m, \\ 
       -1 & \text{otherwise}.
    \end{cases} \label{eq:ideal_kernel}
\end{align}
Although $K^*$ is generally unknown, it can be calculated for the labeled training set.
The estimated matrix $K \in \mathbb{R}^{n_{\text{samples}} \times M}$, on the other hand, contains the fidelity between each datum $\boldsymbol{x}_i$ and each class centroid $\boldsymbol{c}_m$, and is defined later in \cref{eq:mcquack_as_kernel} in \cref{sec:McQuack}.
Given $K$ and $K^*$, their alignment is quantified using the Frobenius inner product
\begin{equation}
\langle A, B\rangle_F = \sum_{i j} A_{i j} B_{i j} = \operatorname{Tr}\left(A^T B\right),
\end{equation}
leading to the kernel alignment measure
\begin{align}
\mathcal{A}(K, K^*) 
= \frac{\langle K, K^*\rangle_F}{\sqrt{\langle K, K\rangle_F \, \langle K^*, K^*\rangle_F}} , \label{eq:kernel_target_alignment}
\end{align}
which is upper bounded by $\frac{1}{\sqrt{M}}$ and where a higher value indicates a stronger alignment between the kernels.
\section{McQuack} \label{sec:McQuack}

In this section, we introduce Multiclass Quantum Aligned Centroid Kernel (\textsc{McQuack}), a \ac{QKM} for multiclass classification with $M$ classes that scales linearly with the training set size and number of classes.
Our model builds upon the Quack algorithm for binary classification \cite{Quack}. However, unlike one-vs-rest or one-vs-one extensions of binary classifiers, our approach uses a single fidelity matrix for all classes.
The model learns a global embedding and class centroids, such that new data can be compared directly to the optimized class centroids in Hilbert space through the fidelity.
For a train set of size $n_\text{train}$, the runtime of the algorithm is in $\mathcal{O}(n_\text{train}\,M)$.
In the following, we first formalize the algorithm, then describe the training process, and finally, map the model to the \ac{VC} formulation.

\subsection{Formalization}
The \textsc{McQuack} classifier is given by
\begin{align}
    F(\boldsymbol{x}) \coloneqq  \bigl[f_0,\,f_1,\,\ldots,\,f_{M-1}\bigr], \label{eq:mcquack_formal}
\end{align}
with $f_m$ defined below. For multiple data points, the model can be formulated as a matrix,
\begin{align}
     K &= \bigl[F(\boldsymbol{x_0}),\,F(\boldsymbol{x_1}),\,\ldots,\,F(\boldsymbol{x_{n_\text{train}-1}})\bigr]^\top \notag \\
    K_{i,m} &= f_m (\boldsymbol{x_i}; \boldsymbol{w}, \boldsymbol{b}, \boldsymbol{c_m}), \quad K \in \mathbb{R}^{n_{\text{train}} \times M} \label{eq:mcquack_as_kernel} 
\end{align}
where each classifier $f_m$ estimates the fidelity between the center of class $m$ and datum $\boldsymbol{x_i}$ as
\begin{align}
    &f_m (\boldsymbol{x_i}; \boldsymbol{w}, \boldsymbol{b}, \boldsymbol{c_m}) = \notag \\ 
    &\vert \bra{0^{\otimes n}} U^\dagger(\boldsymbol{c_m}, \boldsymbol{w}, \boldsymbol{b}) U(\boldsymbol{x_i}; \boldsymbol{w}, \boldsymbol{b}) \ket{0^{\otimes n}} \vert ^2, \label{eq:f_as_kernel}
\end{align}
with shared weights $\boldsymbol{w}$, shared bias $\boldsymbol{b}$ and classifier-specific parameter $\boldsymbol{c_m}$. The unitary $U(\cdot; \boldsymbol{w}, \boldsymbol{b})$ used in our experiments is specified in \cref{sec:appA:McQuack}. The arguments are sometimes omitted for brevity.

\subsection{Training} \label{subsec:training}
\begin{figure*}[t]
\centering

\begin{subfigure}{0.24\textwidth}
    \includegraphics[width=\textwidth]{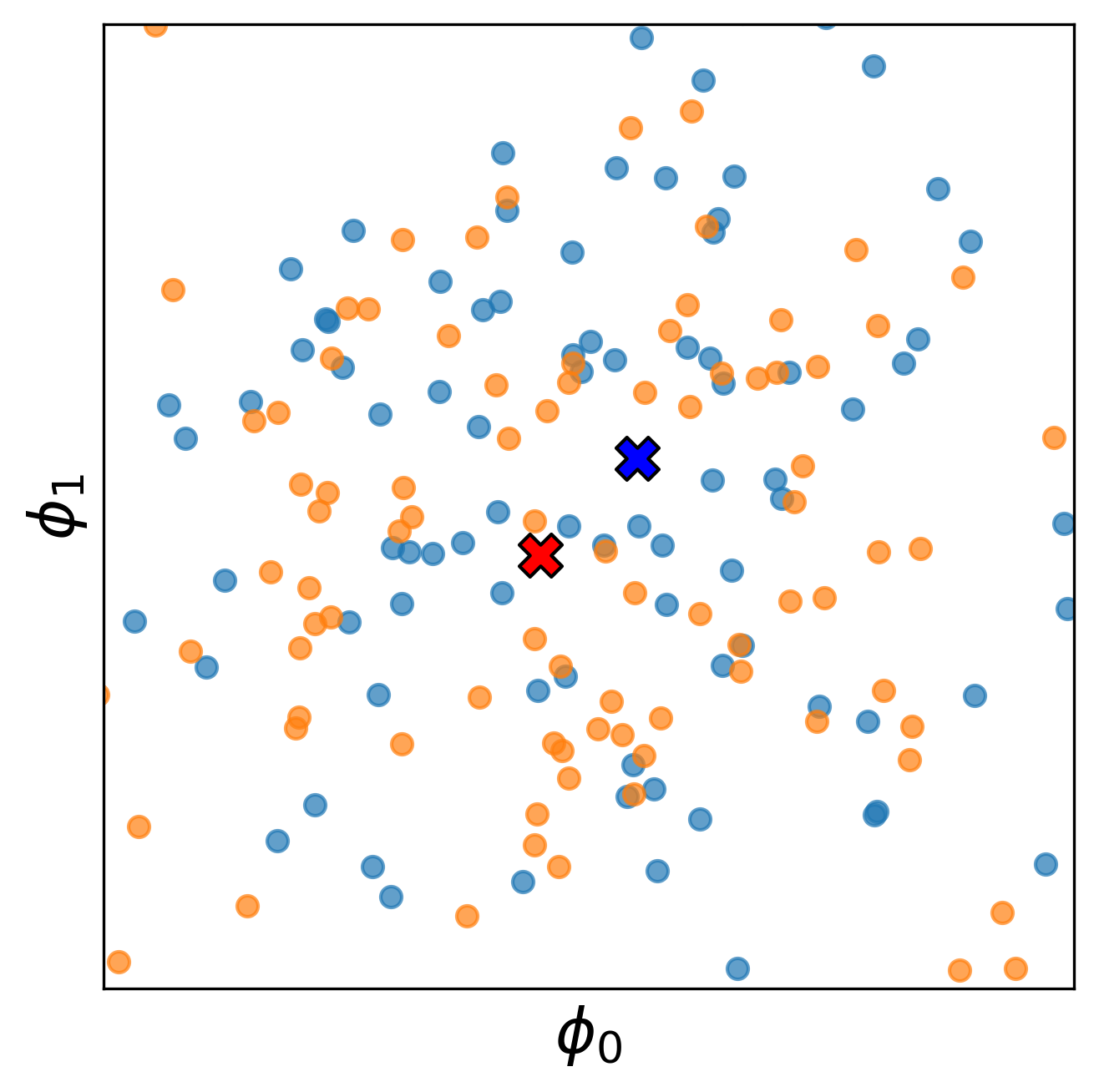}
    \caption{Initial distribution.}
    \label{fig:first}
\end{subfigure}
\hfill
\begin{subfigure}{0.24\textwidth}
    \includegraphics[width=\textwidth]{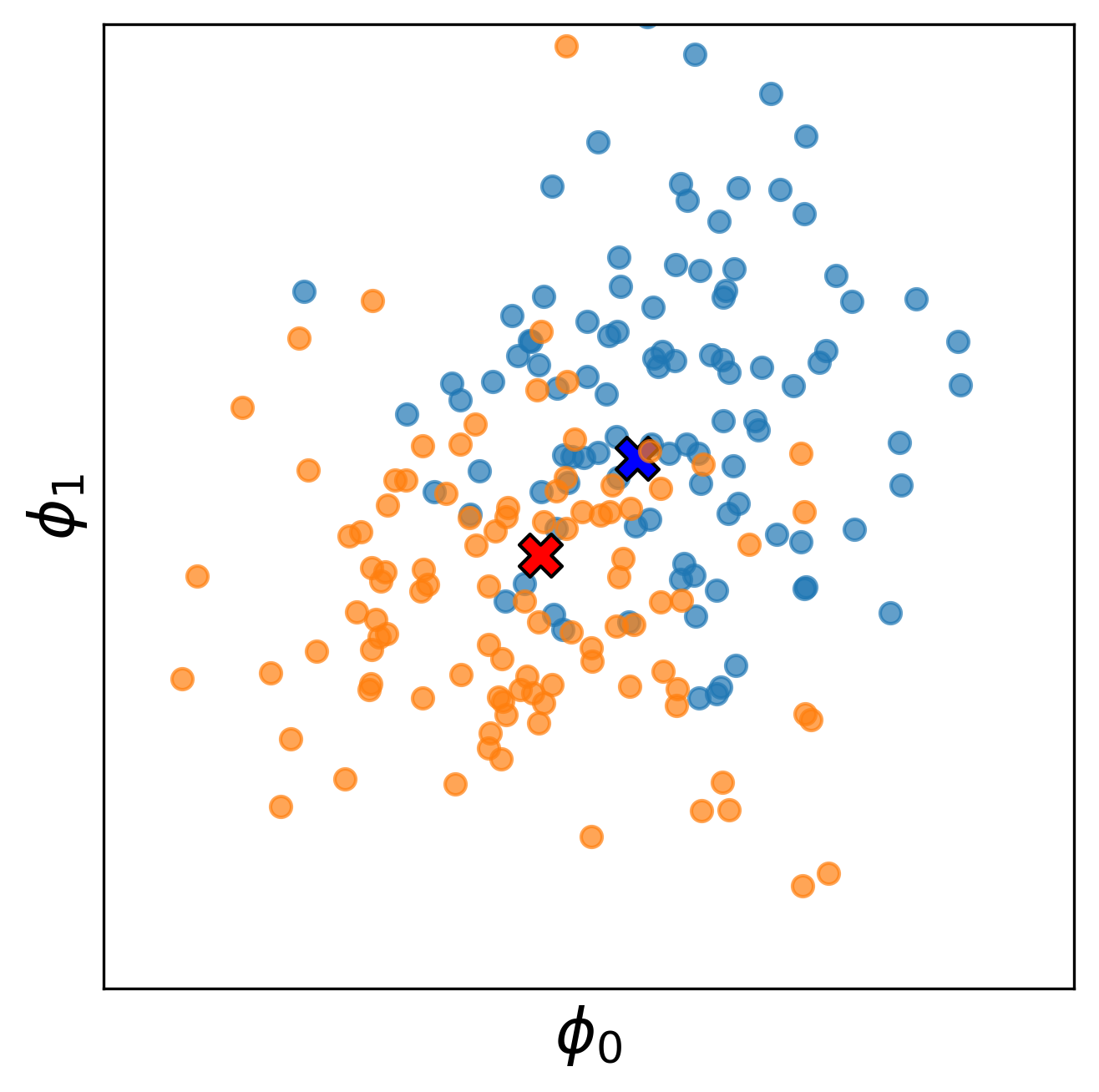}
    \caption{After first KAO step.}
    \label{fig:second}
\end{subfigure}
\hfill
\begin{subfigure}{0.24\textwidth}
    \includegraphics[width=\textwidth]{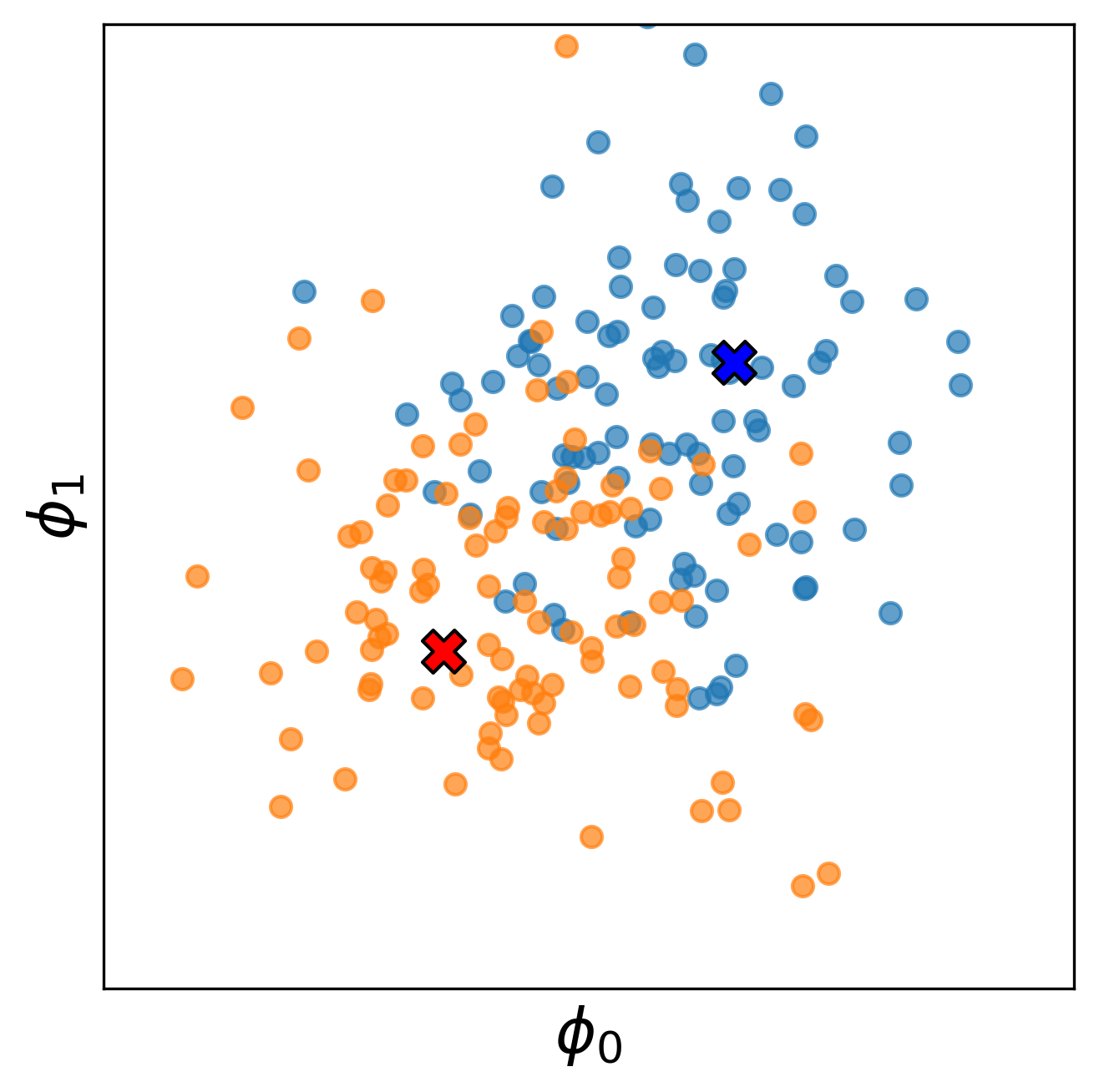}
    \caption{After first CO step.}
    \label{fig:third}
\end{subfigure}
\hfill
\begin{subfigure}{0.24\textwidth}
    \includegraphics[width=\textwidth]{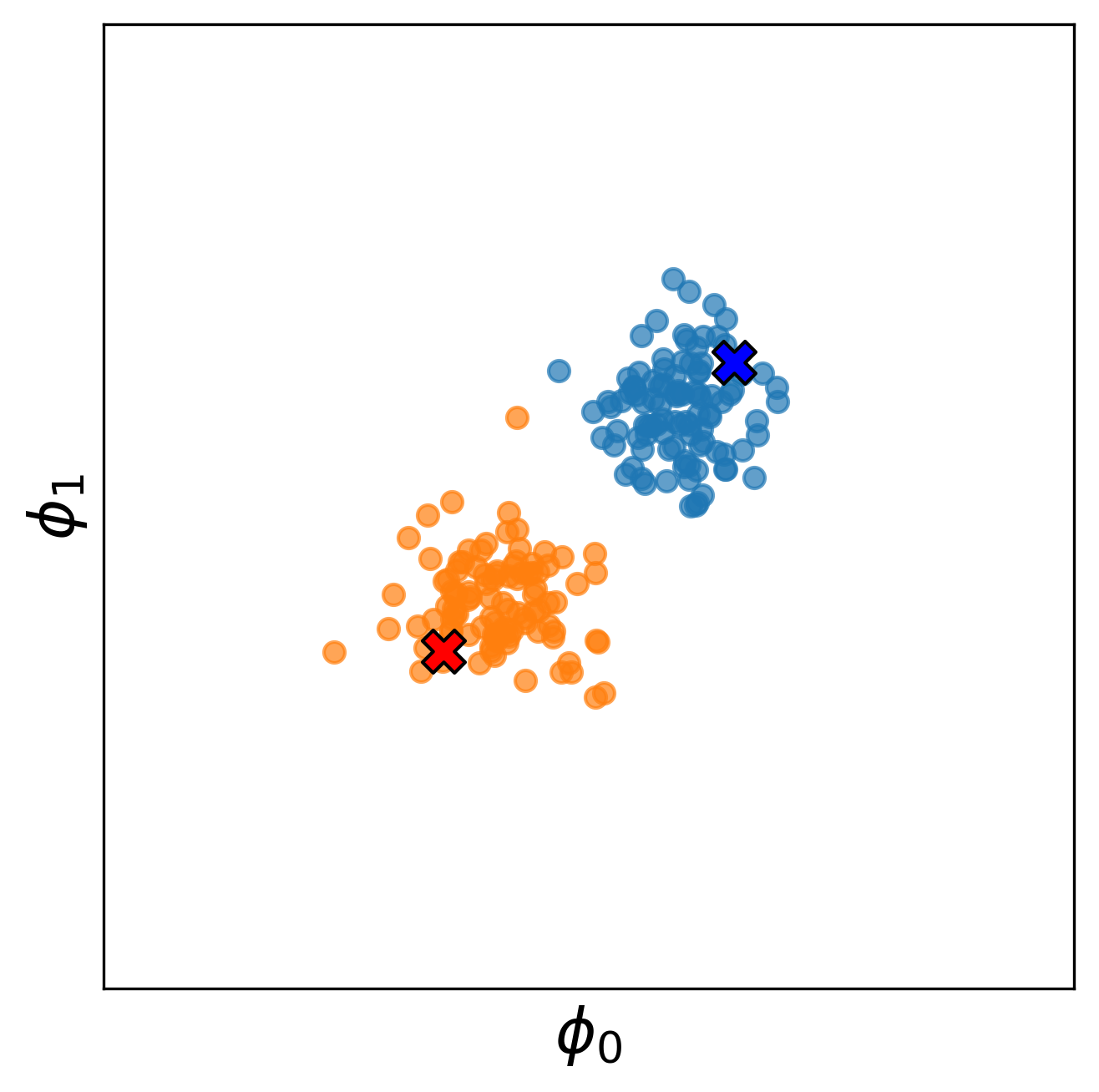}
    \caption{After second KAO step.}
    \label{fig:fourth}
\end{subfigure}

\vspace{0.5em}

\begin{subfigure}{0.24\textwidth}
    \includegraphics[width=\textwidth]{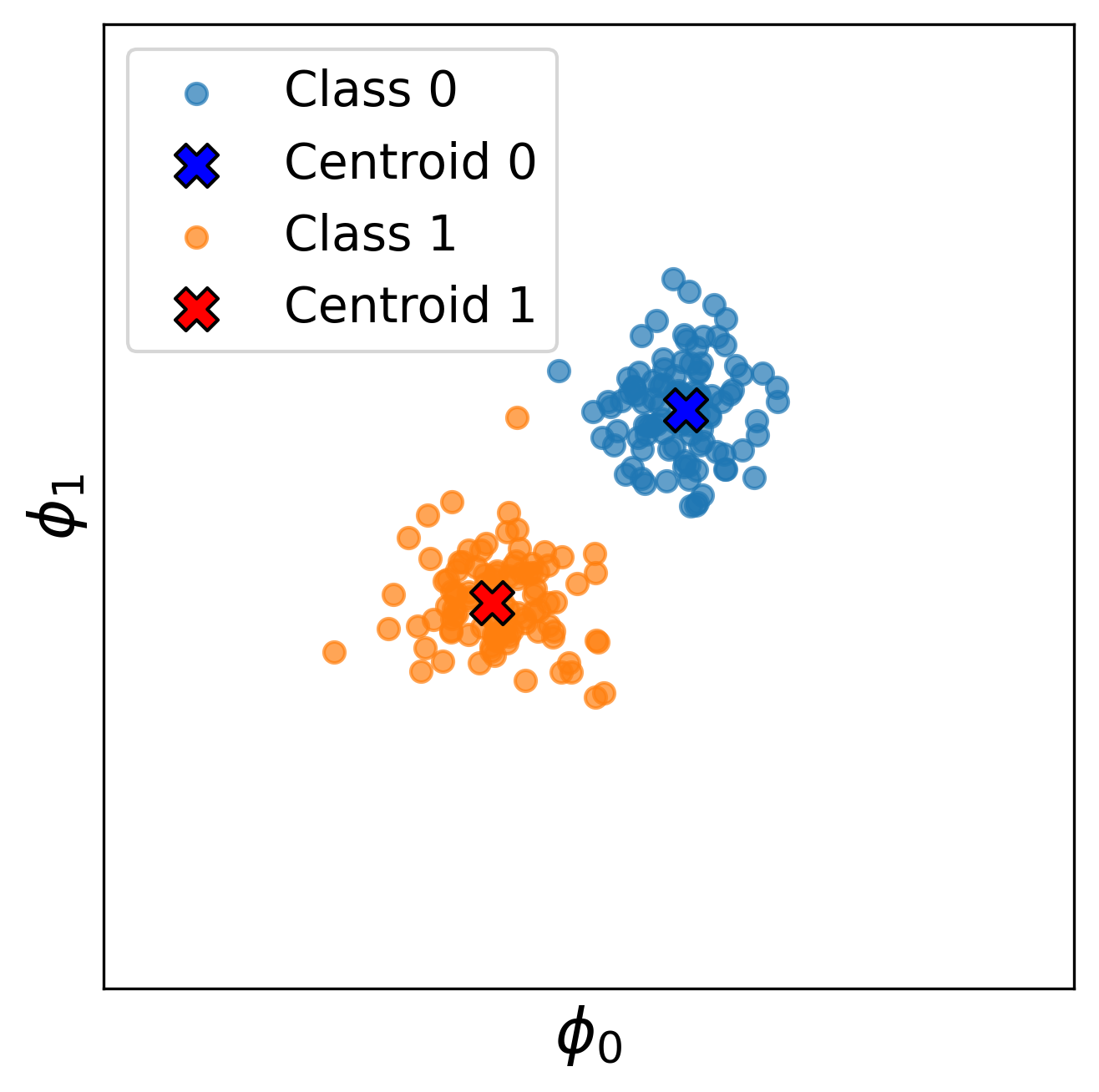}
    \caption{After second CO step.}
    \label{fig:fifth}
\end{subfigure}
\hfill
\begin{subfigure}{0.25\textwidth}
    \caption{Initial kernel.}
    \centering
    \begin{minipage}{\linewidth}

\begin{tabular}{lrrrr}
\toprule
 & $\mathbf{c_0}$ & $\mathbf{c_{1}}$ & $\mathbf{y'}$ & $\mathbf{y}$\\
\midrule
$x_0$ & \valcell{70}{0.7} & \valcell{50}{0.5} & $0$   & $0$   \\
$x_1$ & \valcell{60}{0.6} & \valcell{50}{0.5} & $0$   & $0$   \\
$x_2$ & \valcell{20}{0.2} & \valcell{30}{0.3} & $1$   & $0$   \\
$x_3$ & \valcell{10}{0.1}  & \valcell{70}{0.7} & $1$   & $1$ \\
$x_4$ & \valcell{40}{0.4} & \valcell{2}{0.2} & $0$   & $1$ \\
$x_5$ & \valcell{50}{0.5} & \valcell{10}{0.1}  & $0$   & $1$ \\
\bottomrule
\end{tabular}

    \end{minipage}
    \label{fig:tab1}
\end{subfigure}
\hfill
\begin{subfigure}{0.25\textwidth}
    \caption{Final kernel.}
    \centering
    \begin{minipage}{\linewidth}

\begin{tabular}{lrrrr}
\toprule
 & $\mathbf{c_0}$ & $\mathbf{c_{1}}$ & $\mathbf{y'}$ & $\mathbf{y}$\\
\midrule
$x_0$ & \valcell{90}{0.9} & \valcell{30}{0.3} & 0   & 0   \\
$x_1$ & \valcell{80}{0.8} & \valcell{20}{0.2} & 0   & 0   \\
$x_2$ & \valcell{70}{0.7} & \valcell{10}{0.1} & 0   & 0   \\
$x_3$ & \valcell{10}{0.1} & \valcell{90}{0.9} & 1 & 1 \\
$x_4$ & \valcell{20}{0.2} & \valcell{70}{0.7} & 1 & 1 \\
$x_5$ & \valcell{20}{0.2} & \valcell{60}{0.6} & 1 & 1 \\
\bottomrule
\end{tabular}
    \end{minipage}
    \label{fig:tab2}
\end{subfigure}

\caption{Schematic overview of the \textsc{McQuack} training process on a binary dataset.
(A–E) Data distribution in two dimensions ($\phi_0, \phi_1$) of the Hilbert space after sequential steps of model training.
(A) Initially, the embedding distributes data almost randomly, and the centroids do not represent the class centers well.
(B) During the first \ac{KAO} iteration, the embedding is trained to cluster data points more tightly around their centroids.
(C) The first \ac{CO} iteration optimizes the features of the centroids to shift their position in Hilbert space such that the fidelity with samples from their corresponding class is high and the fidelity with samples from other classes low.
(D) The second \ac{KAO} step further trains the feature map to improve the clustering of the data around their centroids.
(E) The second \ac{CO} iteration again optimizes the centroid locations towards their optimum in Hilbert space.
(F–G) Initial and final kernels (colored columns) with predicted labels ($y'$) and true labels ($y$). Green cells show high kernel values, red ones low values. The initial kernel (F) shows nearly random structure, while the model with final kernel (G) accurately predicts the labels after training the feature map and centroid positions.
}
\label{fig:quack_overview}
\end{figure*}

After formally defining the model, we now look at the training process, outlined in \cref{algo:quack} and illustrated in \cref{fig:quack_overview}. It can be summarized as follows:
McQuack estimates a kernel  $K \in \mathbb{R}^{n_{\text{train}} \times M}$, whose columns are the single classifiers $f_m$ containing fidelities between the samples and class centroids.
Initially, the embedding is random and the centroid positions are not in the class centers (\cref{fig:first}), leading to a largely random kernel (\cref{fig:tab1}).
Each of the $n_\text{epochs}$ iterations alternates between:
\begin{itemize}
    \item $n_\text{KAO}$ epochs of \ac{KAO}, optimizing the embedding parameters $ \boldsymbol{w}, \boldsymbol{b}$ (\cref{fig:second,fig:fourth}),
    \item $n_\text{CO}$ epochs of \ac{CO}, optimizing the centroid features \( \boldsymbol{c_m} \) (\cref{fig:third,fig:fifth}).
\end{itemize}
After training is completed, the centroids produce high fidelities with samples from their corresponding class and low fidelity with samples from other classes, as shown in \cref{fig:fifth}.
The final kernel (\cref{fig:tab2}) correctly classifies the samples by assigning each one the index of the centroid with which it achieves the highest fidelity.
The number of total shots for training the model for $n_\text{epochs}$ epochs using parameter shift rule for a training set of size $n_\text{train}$ and $n_\text{shots}$ shots is 
\begin{align}
    N &= 2 \, n_\text{epochs} \, n_\text{train} \, M \,  n_\text{shots} \, (2 \, n_{\text{KAO}} \, D + 3 \, n_{\text{CO}} \, d) \notag \\ 
    &= \mathcal{O}(n_\text{train} \, M), \label{eq:num_shots_mcquack}
\end{align}
where $D = \text{dim}(\boldsymbol{w}) + \text{dim}(\boldsymbol{b})$.
Note that the parameter shift rule requires two shifted circuit evaluations per parametrized gate occurrence \cite{Schuld_parameter_shift_rule}.
During the inference stage, the predicted label for a new sample is
\begin{align}
    \hat{y_i} = \argmax_{m \in \{0,\ldots,M-1\}} f_m (\boldsymbol{x_i}; \boldsymbol{w}, \boldsymbol{b}, \boldsymbol{c_m}).\label{eq:mcquack_prediction1}
\end{align}

The different parts of the algorithm are described in more detail in \cref{sec:appA:McQuack}.

\begin{figure}[htb]
  \centering
  \begin{minipage}{1.0\linewidth}
    \begin{algorithm}[H]
      \caption{McQuack training} \label{algo:quack}
      \input{1_figures/mcquack_algo_no_logo}
    \end{algorithm}
  \end{minipage}
\end{figure}

\subsection{Mapping to Variational Classifiers} \label{subsec:map_to_vc}

Motivated by the duality between \acp{QKM} and \acp{VC} \cite{schuld2021supervisedquantummachinelearning}, we map \textsc{McQuack} from a kernel-based perspective onto the \ac{VC} framework and discuss how data structure may favor either approach.

To express a single classifier $f_m$ from \cref{eq:f_as_kernel} as a \ac{VC}, we define
\begin{equation}
W(\boldsymbol{x_i}; \boldsymbol{c_m}, \boldsymbol{w}, \boldsymbol{b})
\coloneqq U^\dagger(\boldsymbol{c_m}, \boldsymbol{w}, \boldsymbol{b})\, U(\boldsymbol{x_i}; \boldsymbol{w}, \boldsymbol{b}), \label{eq:w_substitution}
\end{equation}
and substitute to obtain
\begin{align}
    f_m &= \left| \bra{0^{\otimes n}} W \ket{0^{\otimes n}} \right|^2 \notag \\
        &= \bra{0^{\otimes n}} W^\dagger \ket{0^{\otimes n}} \bra{0^{\otimes n}} W \ket{0^{\otimes n}} \notag \\
        &= \bra{0^{\otimes n}} W^\dagger \mathcal{M}_0 W \ket{0^{\otimes n}}, \label{eq:f_as_var_classifier}
\end{align}
with $\mathcal{M}_0 = \ket{0^{\otimes n}}\bra{0^{\otimes n}}$. For readability, arguments are omitted in \cref{eq:f_as_var_classifier}.
Back-substituting $W$ yields the explicit circuit representation
\begin{align}
    f_m = &\bra{0^{\otimes n}} U^\dagger(\boldsymbol{x_i}; \boldsymbol{w}, \boldsymbol{b}) 
    U(\boldsymbol{c_m}, \boldsymbol{w}, \boldsymbol{b}) 
    \mathcal{M}_0 \notag \\ 
    &U^\dagger(\boldsymbol{c_m}, \boldsymbol{w}, \boldsymbol{b}) 
    U(\boldsymbol{x_i}; \boldsymbol{w}, \boldsymbol{b}) 
    \ket{0^{\otimes n}}. 
    \label{eq:mcquack_single_classifier_formal}
\end{align}

This formulation naturally suggests a comparison to an ensemble of \acp{VC}, where each classifier is trained independently. Importantly, $f_m$ depends on both shared parameters $(\boldsymbol{w}, \boldsymbol{b})$ and class-specific parameters $\boldsymbol{c_m}$.

Analogous to \cref{eq:mcquack_formal}, an ensemble of \acp{VC} is given by
\begin{align}
    G(\boldsymbol{x}) = \bigl[g_0,\,g_1,\,\ldots,\,g_{M-1}\bigr], \label{eq:ensemble_vc_formal}
\end{align}
with individual classifiers
\begin{align}
    g_m(\boldsymbol{x_i}; \boldsymbol{\theta_m}) = \langle 0 | V(\boldsymbol{x_i}; \boldsymbol{\theta_m})^{\dagger} \mathcal{M} V(\boldsymbol{x_i}; \boldsymbol{\theta_m}) | 0 \rangle, \label{eq:single_vc_formal}
\end{align}
where $V$ encodes $\boldsymbol{x_i}$ and $\mathcal{M}$ is an observable. The prediction of $G$ is analogous to \cref{eq:mcquack_prediction1} with $g_m$ replacing $f_m$.

\Cref{eq:f_as_var_classifier,eq:single_vc_formal} show that both approaches share a common structure: a unitary parametrized by the input and trainable parameters, followed by measurement. The key difference is that \textsc{McQuack} includes parameters $(\boldsymbol{w}, \boldsymbol{b})$ shared across all classifiers and classifier-specific parameters ($\boldsymbol{c_m}$), whereas standard \ac{VC} ensembles use only classifier-specific parameters ($\boldsymbol{\theta_m}$).
This distinction is also reflected in training where \textsc{McQuack} alternates between optimizing the shared parameters and class-specific centroids.

An intermediate approach can be constructed by introducing shared parameters $\boldsymbol{w}$ into the VC unitary, $V(\boldsymbol{x_i}; \boldsymbol{\theta_m}, \boldsymbol{w})$, and optimizing $\boldsymbol{\theta_m}$ and $\boldsymbol{w}$ alternately. In this case, both models become conceptually equivalent, and performance is expected to depend on the data structure. \\
If classes can be well represented by centroids in Hilbert space, \textsc{McQuack} is likely advantageous. Conversely, ensembles of \acp{VC} with partially shared parameters may perform better when such clustering is not feasible. We therefore encourage further study of \ac{QML} architectures with (partially) shared parameters for multiclass classification.

\section{Results and Discussion} \label{Results and Discussion}
We assess the potential of \textsc{McQuack} in three steps: (i) hardware inference without training on 124-qubit subgraphs of two 156-qubit IBM devices, (ii) extensive benchmarking on 133 synthetic and 19 real-world datasets, and (iii) a trainability analysis. Experimental details are provided in \cref{sec:appB:experiments}.

\subsection{Hardware Results} \label{subsec:hardware_results}
Due to hardware constraints and the overhead introduced by the parameter-shift rule, we evaluate an untrained version of \textsc{McQuack} on 124-qubit subgraphs of two 156-qubit IBM devices.
We employ dynamical decoupling \cite{dynamical_decoupling}, measurement twirling \cite{measurement_twirling} and bit flip tolerance \cite{bitflip_tolerance} as error mitigation strategies. 
For  bit flip tolerance, we developed a calibration procedure tailored to our algorithm, which is explained in \cref{app:bitflip_tolerance_calibration}.
The binary dataset follows the manifold hypothesis and is described in \cref{app:hardware_dataset}.
We compare against an \ac{SVC} and an untrained \ac{RBF} kernel (SimpleKernel), the latter chosen to enable a fair comparison with an untrained classical model.
Differences are not statistically significant (for $\alpha = 0.05$) due to limited runs, hence these results should be interpreted as a feasibility demonstration of inference on large-scale hardware rather than evidence of practical hardware competitiveness.

The untrained \textsc{McQuack} achieves an accuracy of $0.83 \pm 0.02$ on \texttt{ibm\_pittsburgh} and $0.79 \pm 0.02$ on \texttt{ibm\_aachen}, compared to $0.86$ for the \ac{SVC} and $0.80$ for \textsc{SimpleKernel}, as summarized in \cref{tab:hardware_results}. The trend toward improved performance on \texttt{ibm\_pittsburgh} relative to \texttt{ibm\_aachen} may be attributed to the advanced processor on the former device, leading to lower error rates (see \cref{tab:ibm_aachen_stats} in \cref{subsec:appB:hardware_experiments}). 
Performance of \textsc{McQuack} on \texttt{ibm\_pittsburgh} was in the same range as the untrained RBF baseline, suggesting that fidelity evaluation remains feasible on 100+ qubits.
The performance gap to \ac{SVC} is expected, since \textsc{McQuack} performs no training and scales linearly in the number of training samples, in contrast to the quadratic scaling of the \ac{SVC}. 

Training on hardware is currently infeasible due to the high cost of gradient estimation with the parameter shift rule, resulting in prohibitively long runtimes. 
Optimizing the model for one epoch consisting of one iteration of \ac{KAO} and one of \ac{CO} for the dataset and architecture from the hardware experiments would require $\approx 1.5 \times 10^8$ shots according to \cref{eq:num_shots_mcquack}. Considering the execution time for our experiments of approximately two seconds per 100 shots, a single epoch of \textsc{McQuack} training would take $\approx 827\,\text{h}$ or over 34 days of QPU time. 
We therefore focus on simulation-based evaluation for the remaining experiments.

\begin{table}[ht]
\centering
\caption{Results of the untrained \textsc{McQuack} model on hardware (Pittsburgh and Aachen) and the classical benchmarks (SVC, SimpleKernel) in descending order. Statistics are collected from three runs of each model and dataset.}
\label{tab:hardware_results}
\begin{tabular}{lll}
\toprule
Model & Accuracy & Acc. Values \\
\midrule
SVC & 0.86 & 0.86, 0.86, 0.86 \\
Pittsburgh & 0.83 $\pm$ 0.02 & 0.86, 0.84, 0.80 \\
SimpleKernel & 0.80 & 0.80, 0.80, 0.80 \\
Aachen & 0.79 $\pm$ 0.02 & 0.80, 0.80, 0.76 \\
\bottomrule
\end{tabular}

\end{table}

\subsection{Benchmark}
To evaluate \textsc{McQuack}, we perform a benchmark on 133 synthetic datasets split into five classification tasks from the QML Benchmark Suite and on 19 real-world datasets.

\subsubsection{QML Benchmark Suite}
\begin{figure}[htbp]
\centering

\begin{subfigure}[t]{0.48\textwidth}
    \centering
    \includegraphics[width=\textwidth]{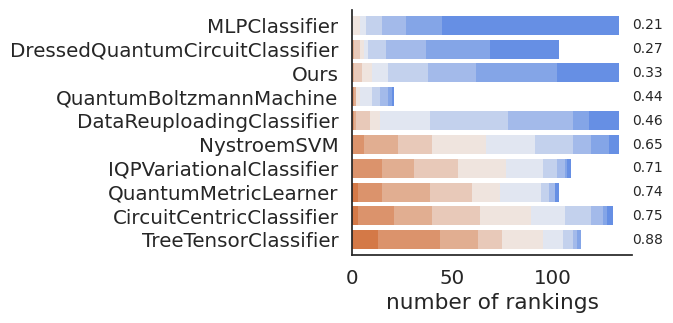}
    \caption{Comparison of model rankings by accuracy for the quantum neural net family of the \ac{QML} Benchmark. Some models are not evaluated on all datasets and consequently have fewer rankings. The out-of-the-box classical \textsc{MLPClassifier} is superior to all quantum models. \textsc{McQuack} is the second best quantum model, being only outperformed by the \textsc{DressedQuantumCircuitClassifier}, which uses classical neural nets for pre- and postprocessing.}
    \label{fig:ranking_qml_benchmarks}
\end{subfigure}
\hfill
\begin{subfigure}[t]{0.48\textwidth}
    \centering
    \includegraphics[width=\textwidth]{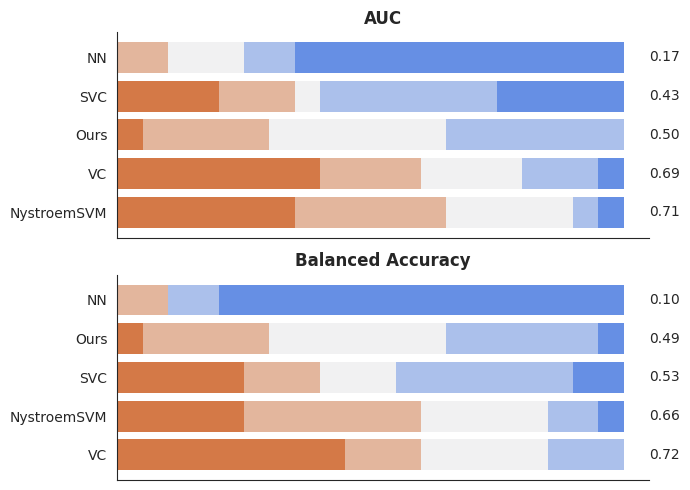}
    \caption{Comparison of model rankings by AUC and balanced accuracy for the real-world datasets. The neural network performs best while \textsc{McQuack} outperforms the other quantum model (\ac{VC}) and the efficient classical kernel method (\textsc{NystroemSVM}).}
    \label{fig:ranking_real_world_ds}
\end{subfigure}

\caption{Number of rankings (blue/first to red/last) for the QML Benchmark suite (left) and real-world datasets (right). The models are sorted from top to bottom according to the average normalized rank, which is listed on the right. The plotting style is adapted from \cite{schuld_qml_benchmarks2024}.}
\label{fig:ranking_combined}

\end{figure}

We compare \textsc{McQuack} against the quantum neural network family from the \ac{QML} Benchmark Suite \cite{schuld_qml_benchmarks2024}, consisting of seven quantum models, and the \textsc{MLPClassifier}\footnote{\url{https://scikit-learn.org/stable/modules/generated/sklearn.neural_network.MLPClassifier.html}} as a classical baseline.
Additionally, we report results for \textsc{NystroemSVM}, a linear SVM that uses the Nystroem method to approximate an RBF kernel. This method utilizes six landmark points to approximate the kernel, which results in a probability of approximately $97\%$ that the kernel includes at least one landmark point from each class.
This baseline allows us to compare \textsc{McQuack} with a classical model that also uses a kernel approximation.
Although limiting \textsc{NystroemSVM} to six landmark points reduces its number of trainable parameters, it still uses more landmark points than \textsc{McQuack} has centroids to approximate the kernel. However, unlike the randomly sampled landmark points of \textsc{NystroemSVM}, the two centroids that \textsc{McQuack} utilizes are trainable.
\Cref{fig:ranking_qml_benchmarks} plots the normalized rankings of the models, where a lower average rank is better.
As reported in \cite{schuld_qml_benchmarks2024}, the classical \textsc{MLPClassifier} achieves the best performance, with an average rank of 0.21.
\textsc{NystroemSVM}, the second classical baseline, ranks near  the middle with an average rank of 0.65. This may suggest that using a small number of random landmark points to approximate the RBF kernel leads to a less effective representation of the data than utilizing trainable centroids in a Hilbert space, as \textsc{McQuack} does.
With an average normalized rank of 0.33, our model ranks third-best, just behind the \textsc{DressedQuantumCircuitClassifier} \cite{DressedQuantumCircuitClassifier} whose average rank is 0.27.
However, the latter is a hybrid model that uses trainable neural networks for pre- and post-processing, thus preventing a fair comparison. Furthermore, it has not been evaluated on all 133 datasets. \\
Our model ranks better than the remaining quantum models, including the \textsc{DataReuploadingClassifier} \cite{reupload_encoding_2020}, which uses a re-uploading encoding similar to ours. 
Detailed results in \cref{subsec:appC:qml_benchmark_suite} show consistent rankings and no strong dependence on dataset dimensionality. We therefore identify \textsc{McQuack} as the strongest "pure" quantum model\footnote{Although our model is still hybrid in the sense that it uses a classical optimizer, like most current \ac{QML} models.} in this benchmark.

\subsubsection{Real-World Datasets}
The performance of \textsc{McQuack} was evaluated on 19 real-world datasets from two to ten classes ranging over a variety of use cases with the goal of providing insights into the model's applicability in practical scenarios. Additionally, the manifolds dataset from \cref{subsec:hardware_results} is included in this benchmark. \\
We compare our model against one quantum and three classical baselines: an ensemble of \acp{VC} as described in \cref{eq:ensemble_vc_formal,eq:single_vc_formal} as quantum reference, a \ac{SVC} as powerful kernel-based baseline, the \textsc{NystroemSVM} with $3M$ landmark points as efficient kernel-based method, and a two-layer \ac{NN} with a comparable number of trainable parameters as \textsc{McQuack}.
The aggregated model rankings are presented in \cref{fig:ranking_real_world_ds}, and the detailed numerical results are reported in \cref{tab:results_real_world_ds} in \cref{app:real_world_datasets}. 
The \ac{NN} demonstrates superior performance, attaining the best overall rank of 0.17 for AUC and 0.10 for balanced accuracy, aligning with the observation from the previous section.
\textsc{McQuack}, with mean ranks of 0.50 regarding AUC and 0.49 regarding balanced accuracy, is competitive with the remaining models: it ranks below the SVC in terms of AUC but better than the SVC regarding the balanced accuracy, despite its linear scaling (vs.\ quadratic of the \ac{SVC}).
Moreover, \textsc{McQuack} ranks better than \textsc{NystroemSVM} which is an efficient classical kernel method.
In addition, our model outperforms the ensemble of \acp{VC}, demonstrating that the use of partially shared trainable parameters yields a competitive advantage over purely individual ones.

\subsection{Trainability Study}

This section investigates the impact of regularization and initialization on training dynamics. 
The use of regularization is motivated by its proven effectiveness in reducing overfitting in ML \cite{deep_learning_book_Goodfellow} and improving model robustness in \ac{QML} \cite{qml_robustness_lipshitz, wendlinger_adv_robustness}.
For parameter initialization, we consider Gaussian initialization with zero mean and small standard deviation; initialization with ones; and an additional strategy that on average enforces zero angles for the first batch, i.e., $w \cdot x^{(d)} + b = 0$, where $x^{(d)}$ is the batch-wise mean of the $d$-th feature (see \cref{subsec:experiments_mcquack}).

\begin{figure}[htb]
\centering
    \includegraphics[width=0.5\textwidth]{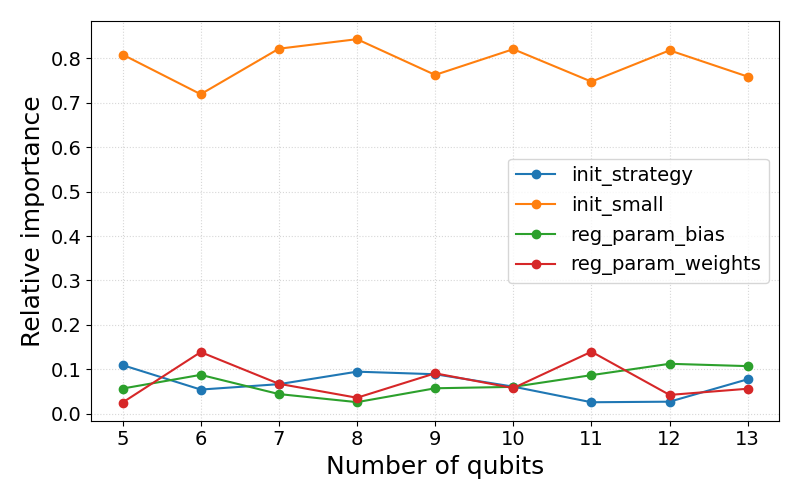}
\caption{fANOVA analysis showing the relative importance of individual hyperparameters.}
\label{fig:hyperparam_importance_fanova}
\end{figure}

\begin{figure*}[htb]
\centering
\includegraphics[width=\textwidth]{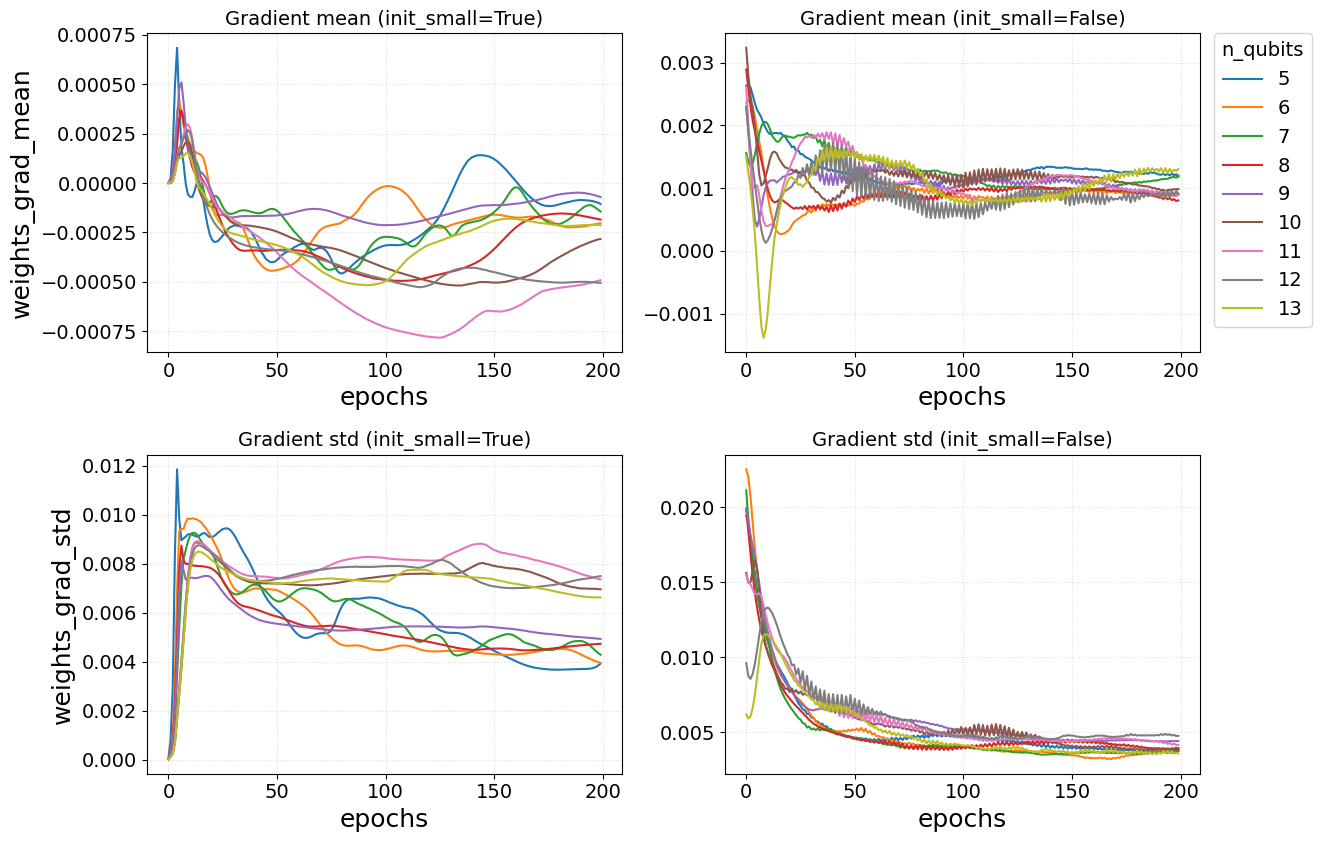}
\caption{ Mean and standard deviation of the weight gradients computed for the initial batch of the \ac{KAO} routine on the manifold dataset for the regularized model with wx+b=0 initialization strategy. Note the different scales on the y-axis.}
\label{fig:trainability_grads_mean_std}
\end{figure*}

\begin{figure}[t]
    \centering
    \includegraphics[width=0.48\textwidth]{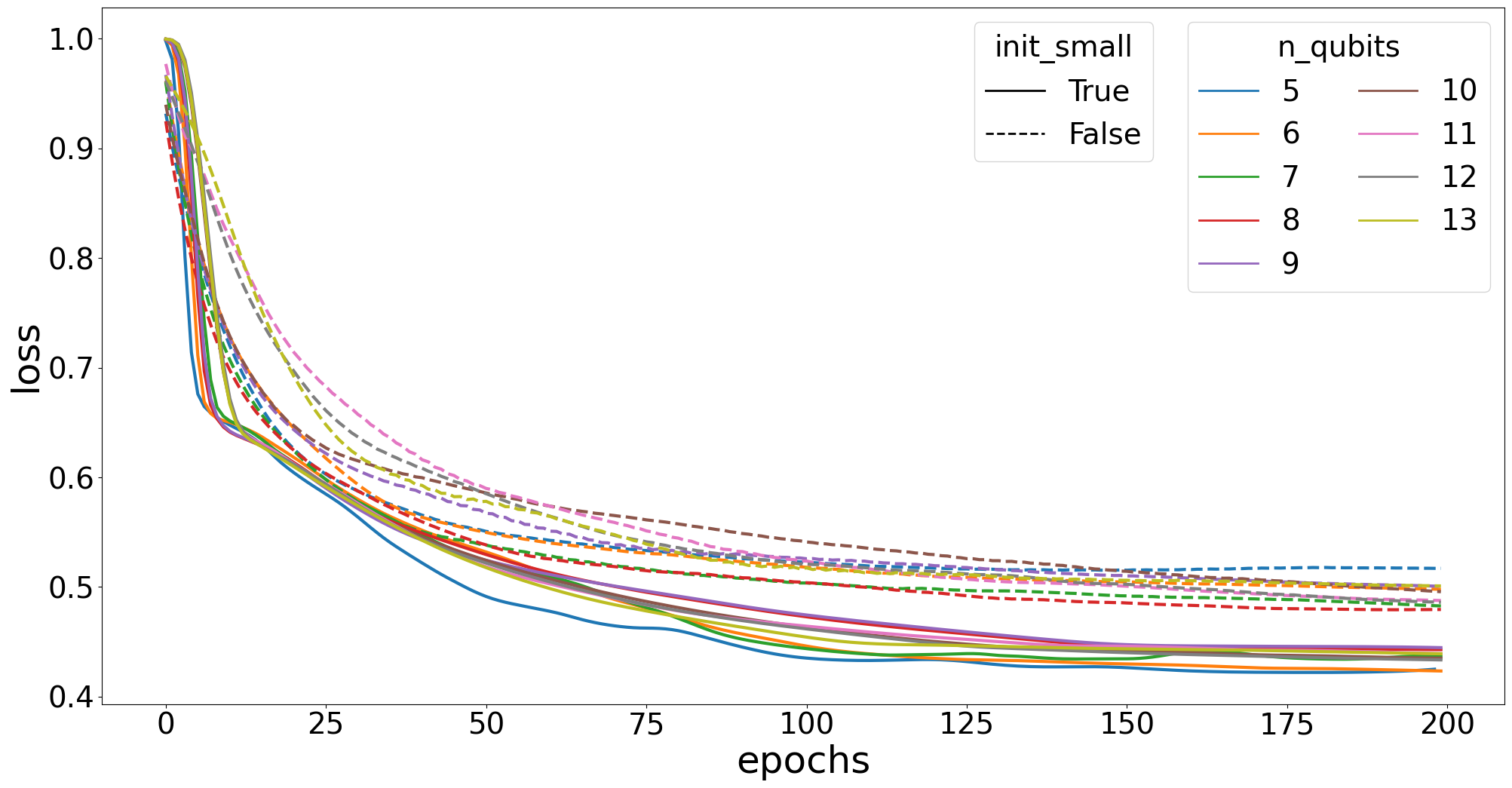}
    \caption{Training loss for different system sizes and magnitudes of the initial parameters on the manifold dataset for the regularized model with wx+b=0 initialization strategy.}
    \label{fig:trainability_loss}
\end{figure}

The fANOVA analysis (\cref{fig:hyperparam_importance_fanova}) identifies initialization scale and distribution (\texttt{init\_small}) as the dominant factor, with relative importance exceeding 0.7, while all remaining hyperparameters contribute less than 0.2. \\
Models initialized with small, normally distributed parameters exhibit stable gradients and improved convergence, while large initializations lead to oscillatory behavior (\cref{fig:trainability_grads_mean_std,fig:trainability_loss}). These trends are consistent across system sizes and reflected in both gradient statistics and training loss.

Overall, we observe no evidence of barren plateaus up to 13 qubits. 
While both the magnitude and distribution of the initial parameters strongly affect model performance, disentangling their individual contributions remains an open question for future work.

\section{Conclusion and Outlook} \label{Conclusion and Outlook}
This work uses \ac{QML} to address three fundamental limitations of kernel methods in machine learning: quadratic scaling with the training set size, reliance on fixed and non-trainable kernels, and the lack of an intrinsic formulation for multiclass classification.
We have accomplished this by introducing \textsc{McQuack}, a multiclass \ac{QKM} based on a trainable kernel that achieves linear runtime complexity with respect to the training set size and the number of classes.
We assessed the potential of the model in three settings.
First, evaluating the untrained model on 124 qubits across two IBM systems revealed no statistically significant difference in accuracy relative to the baselines. Moreover, in one run, our model on \texttt{ibm\_pittsburgh} matched the \ac{SVC}, despite having no trained parameters and a lower runtime complexity. On that device, \textsc{McQuack} was also at least as effective as the \ac{RBF} kernel in all runs. \\
Second, benchmarking against the quantum neural network family of the \ac{QML} Benchmark Suite demonstrated that \textsc{McQuack} consistently outperforms existing "pure" quantum models.
Additionally, \textsc{McQuack} ranks above a linear SVM with Nystroem approximation of the RBF kernel, suggesting that trainable centroids mapped to a Hilbert space provide a more effective representation than a small number of random RBF landmark points.
However, \textsc{McQuack} remains inferior to the other hybrid and fully classical approaches.
In an additional benchmark on 19 real-world datasets, \textsc{McQuack} outperformed a \acl{VC} with the same architecture, and achieved performance comparable to an \ac{SVC}, despite the latter’s quadratic runtime complexity. As in the synthetic benchmark, \textsc{McQuack} ranked better than the \textsc{NystroemSVM} but classical neural networks demonstrated the highest overall performance. \\
Finally, a trainability study revealed no evidence of barren plateaus for system sizes up to 13 qubits. We further showed that the magnitude and distribution of the initial parameters strongly influence model performance, whereas the effects of regularization and identity-based initialization are less pronounced.

Taken together, these results indicate that \textsc{McQuack} is a promising approach for multiclass classification, especially once advances in quantum hardware and software reduce noise levels and enable efficient parameter optimization on hardware.
As discussed in \cref{subsec:map_to_vc}, we encourage further investigation into ensembles of \acp{VC} with partially shared parameters, and assessment of their potential for multiclass classification.
Further research directions include systematic comparisons between our model's kernel and those of other \acp{QKM}, deeper studies of parameter initialization and its impact on optimization behavior, and evaluating the model's robustness against adversarial perturbations to determine its applicability in security-sensitive settings.

\section{Limitations} \label{sec:limitations}

Despite the promising results, particularly when compared to other \ac{QML} models, several limitations remain.
First, training on quantum hardware is currently infeasible for practically relevant system sizes due to the high cost of gradient estimation via the parameter-shift rule. As a result, hardware experiments are restricted to untrained models, limiting conclusions about practical performance. Moreover, the simulations are performed on a noiseless simulator, which may overestimate performance compared to noisy hardware. 
Second, while \textsc{McQuack} is competitive among quantum models, it does not outperform simple classical baselines like feedforward neural networks.
Third, the trainability analysis is limited to system sizes of up to 13 qubits. Although no barren plateaus are observed in this regime, these results may not generalize to larger-scale quantum systems where optimization landscapes can differ drastically. 
Finally, the method implicitly assumes that classes can be represented by centroids in Hilbert space, which may not hold for more complex data distributions.

\section{Code Availability}
Code for the hardware experiments and real-world benchmarks is pending export control clearance and will be made publicly available at a later stage. In the meantime, the source code can be obtained from the authors upon reasonable request.
Code for the QML Benchmark Suite is adapted from \cite{schuld_qml_benchmarks2024} and should be taken from there.

\section*{Acknowledgment}
This research is part of the Munich Quantum Valley (MQV), which is supported by the Bavarian state government with funds from the Hightech Agenda Bayern Plus.

\section*{Author Contribution Statements}
KT carried out the research and wrote the manuscript. PD supervised the project and strongly influenced it with his ideas. AI was used solely for grammar and spelling checks.

\acrodef{ANOVA}{Analysis of Variance}
\acrodef{CO}{Centroid Optimization}
\acrodef{KAO}{Kernel Alignment Optimization}
\acrodef{KM}{Kernel Method}
\acrodef{KTA}{Kernel-Target Alignment}
\acrodef{NN}{Neural Network}
\acrodef{QC}{Quantum Computing}
\acrodef{QKM}{Quantum Kernel Method}
\acrodef{QML}{Quantum Machine Learning}
\acrodef{RBF}{Radial Basis Function}
\acrodef{SVC}{Support Vector Classifier}
\acrodef{VC}{Variational Classifier}


\bibliographystyle{quantum}
\bibliography{bib}

\clearpage
\onecolumn
\appendix

\section{McQuack} \label{sec:appA:McQuack}
This appendix describes the different parts of the Multiclass Quantum Aligned Centroid Kernel (\textsc{McQuack}) algorithm in detail.

\subsection{Circuit Design and Data Encoding} \label{Circuit Design and Data Encoding}
We follow the trend of using hardware-efficient, trainable encodings which were found to yield robustness and generalization improvements over fixed ones \cite{qml_robustness_lipshitz}. 
The trainable feature map $U(\cdot; \boldsymbol{w}, \boldsymbol{b})$ consists of $L$ layers, and each layer $U_l$ is composed of a rotation unitary $U_\text{ROT}(\boldsymbol{x_i}; \boldsymbol{w_l}, \boldsymbol{b_l})$ and an entangling unitary $U_\text{ENT}$,
\begin{align}
    U\left(\boldsymbol{x_i}; \boldsymbol{w}, \boldsymbol{b}\right) = \prod_{l=0}^{L-1} U_l \left(\boldsymbol{x_i}; \boldsymbol{w_l}, \boldsymbol{b_l}\right)  
    = \prod_{l=0}^{L-1} U_\text{ENT} U_{\text{ROT}} \left(\boldsymbol{x_i}; \boldsymbol{w_l}, \boldsymbol{b_l}\right). \label{eq:full_U_definition}
\end{align}

\begin{figure}[htb]
\centerline{\includegraphics[width=.5\textwidth]{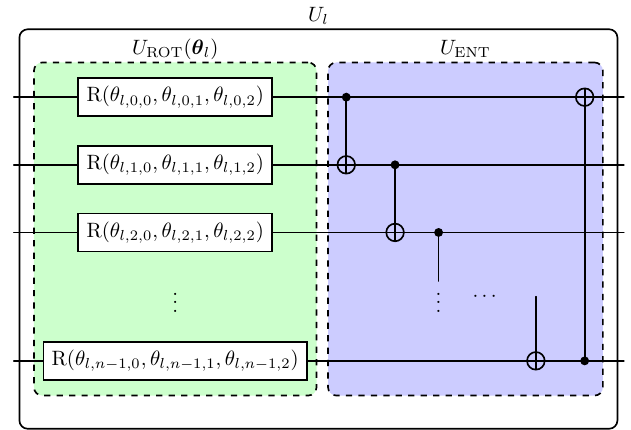}}
\caption{Circuit of a layer unitary $U_l$ consisting of the rotation unitary (green) and the entangling unitary (blue).}
\label{fig:one_layer}
\end{figure}

\begin{figure}[htb]
\centerline{\includegraphics[width=1.0\textwidth]{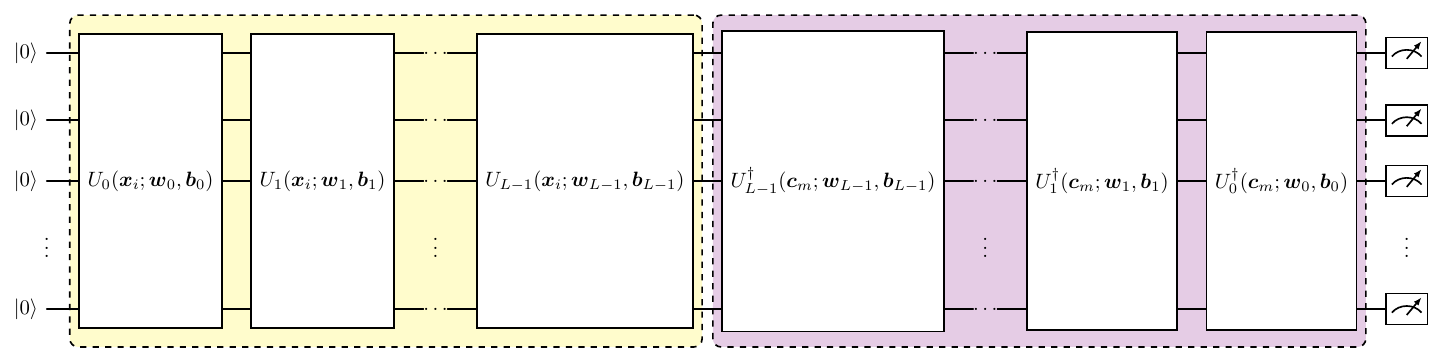}}
\caption{Schematic of the full \textsc{McQuack} circuit. The yellow section encodes the sample $\boldsymbol{x_i}$ and the purple section the centroid $\boldsymbol{c_m}$. A single layer unitary $U_l$ is shown in \cref{fig:one_layer}.}
\label{fig:full_circuit}
\end{figure}

A single layer unitary $U_l$ is sketched in \cref{fig:one_layer} and the full circuit is shown in \cref{fig:full_circuit}.
The rotational unitary is constructed from the general single-qubit rotation gate $R$ \cite{Quantum_Computing_Schuld2021} as
\begin{align} \label{eq:U_ROT}
    U_{\text{ROT}} \left(\boldsymbol{x_i}; \boldsymbol{w_l}, \boldsymbol{b_l}\right) = \bigotimes_{q=0}^{n-1} R\left(\theta_{l,q,0}, \,\theta_{l,q,1}, \, \theta_{l,q,2}\right),
\end{align}
with
\begin{align}
&R\left(\theta_{l,q,0}, \, \theta_{l,q,1}, \, \theta_{l,q,2}\right) = R(\phi, \theta, \omega) 
=R Z(\omega) R Y(\theta) R Z(\phi) 
= \left[\begin{array}{cc}
e^{-i(\phi+\omega) / 2} \cos (\theta / 2) & -e^{i(\phi-\omega) / 2} \sin (\theta / 2) \\
e^{-i(\phi-\omega) / 2} \sin (\theta / 2) & e^{i(\phi+\omega) / 2} \cos (\theta / 2)
\end{array}\right], \label{eq:Rot_gate} 
\end{align}
where for simplicity, the angles are substituted by
$
    \phi \coloneqq \theta_{l,q,0}, \; \theta \coloneqq \theta_{l,q,1}, \; \omega \coloneqq \theta_{l,q,2}.
$
The parameters of $R$ in \cref{eq:U_ROT,eq:Rot_gate} are defined analogously to a neuron in a neural network:
\begin{align}
    \theta_{l,q,0} &= w_{l,q,0} \cdot x_{i, ln+q\text{ mod } d} + b_{l,q,0} \notag \\
    \theta_{l,q,1} &= w_{l,q,1} \cdot x_{i, ln+q\text{ mod } d} + b_{l,q,1} \notag \\
    \theta_{l,q,2} &= w_{l,q,2} \cdot x_{i, ln+q\text{ mod } d} + b_{l,q,2}
\end{align}
for the $i$-th datum with dimension $d$. In other words, a single feature is encoded in all three angles of the $R$-gate. 
If the number of rotational gates exceeds the data dimension, a re-uploading encoding \cite{reupload_encoding_2020} is used, where the gates re-encode the features cyclically from the beginning.

The entangling unitary is a ring of $CNOT$s, connecting each qubit to the next one and the last one to the first one,
\begin{align}
   U_{\text{ENT}} = \prod_{q = 0}^{n-1} CNOT_{q \rightarrow (q+1)\text{ mod } n}. 
\end{align}

\subsection{Kernel Alignment Optimization} 
\label{Kernel Alignment Optimization}
During the \acl{KAO}, the parameter vectors $\boldsymbol{w}$ and $\boldsymbol{b}$ of the feature map are optimized. 
The loss $L_\text{KAO}$ is calculated from the \acl{KTA} $\mathcal{A}(K, K^*)$ (see \cref{eq:kernel_target_alignment}) with $K$ obtained from \cref{eq:mcquack_as_kernel} and $K^*$ from \cref{eq:ideal_kernel}.
Regularization is performed via the parameters $\lambda_w, \lambda_b$. Note that during this step, the centroids $\boldsymbol{c_m}$ are fixed.
\begin{align}
L_\text{KAO} (C; \boldsymbol{w}, \boldsymbol{b}) = 1 - \mathcal{A}(K, K^*)  + \lambda_{w} ||\boldsymbol{w}||^2_2 + \lambda_b ||\boldsymbol{b}||^2_2
\end{align}
The term $1 - \mathcal{A}(K, K^*)$ ensures that the loss is non-negative\footnote{if the lower bound of the loss is desired to be zero, a term of $\frac{1}{\sqrt{M}}$ should be used instead of one.}.
This loss function is then used to optimize the kernel parameters $\boldsymbol{w}$ and $\boldsymbol{b}$ either through backpropagation (in simulations) or the parameter shift rule by solving the minimization problem: 
$
    \min_{\boldsymbol{w}, \boldsymbol{b}} L_\text{KAO} (C; \boldsymbol{w}, \boldsymbol{b})
$

\subsection{Centroid Optimization}
The \acl{CO} optimizes the centroids $C = \bigl[\boldsymbol{c_0},\,\boldsymbol{c_1},\,\ldots,\,\boldsymbol{c_{M-1}}\bigr]$ in data space. For this, the \ac{KTA} is calculated the same way it is in the \ac{KAO} routine and then translated into a loss function $\mathcal{L}_\text{CO}$, in which the parameters $\boldsymbol{w}$ and $\boldsymbol{b}$ of the feature map are fixed:
\begin{align}
L_\text{CO}(\boldsymbol{w}, \boldsymbol{b}; C) = 1 - \mathcal{A}(K, K^*) + \lambda_{\text{CO}} R
\end{align}
Since the features are normalized, the regularization term $R$ penalizes centroid positions that lie outside the normalization range:
\begin{align}
R &= \sum_{m=0}^{M-1}\sum_{f=0}^{d-1} \left(\max(\boldsymbol{c}^f_{m}-1  ,0) - \min(\boldsymbol{c}^f_{m}, 0)\right),
\end{align}
where $\boldsymbol{c}^f_{m}$ is the $f$-th feature of the centroid $\boldsymbol{c}_{m}$. The loss is then minimized accordingly:
$
    \min_{C} L_\text{CO} (\boldsymbol{w}, \boldsymbol{b}; C) 
$

\subsection{Inference}
Once the model is trained, the predicted class of a new sample $\boldsymbol{x_i}$ is the index of the classifier with the highest output,
\begin{align}
    \hat{y_i} = \argmax_{m \in \{0,\ldots,M-1\}} f_m(\boldsymbol{x_i}), 
\end{align}
The runtime complexity of the model for $n_\text{test}$ samples in the inference stage is in $\mathcal{O}(n_\text{test}M)$.

\clearpage
\section{Experiments} \label{sec:appB:experiments}

In order to assess the potential of \textsc{McQuack}, we (a) evaluate the model on 124 qubits of IBM hardware, (b) carry out an extensive and rigorous benchmark on 133 synthetic and 19 real world datasets, and (c) study the trainability of the model.
The details of the experiments are described in the following.
The source code of the model will be made available in a public repository upon publication.

\subsection{Hardware Experiments} \label{subsec:appB:hardware_experiments}
A first assessment of the model's performance on hardware is obtained by estimating a kernel on 156-qubit IBM Pittsburgh and Aachen, which are characterized in \cref{tab:ibm_aachen_stats}. 
Dynamical decoupling \cite{dynamical_decoupling}, measurement twirling \cite{measurement_twirling} and bit flip tolerance \cite{bitflip_tolerance} are used as error mitigation techniques. The calibration procedure for bit flip tolerance tailored to our algorithm is described in \cref{app:bitflip_tolerance_calibration}.

\begin{figure}[htb]
\centerline{\includegraphics[width=0.8\textwidth]{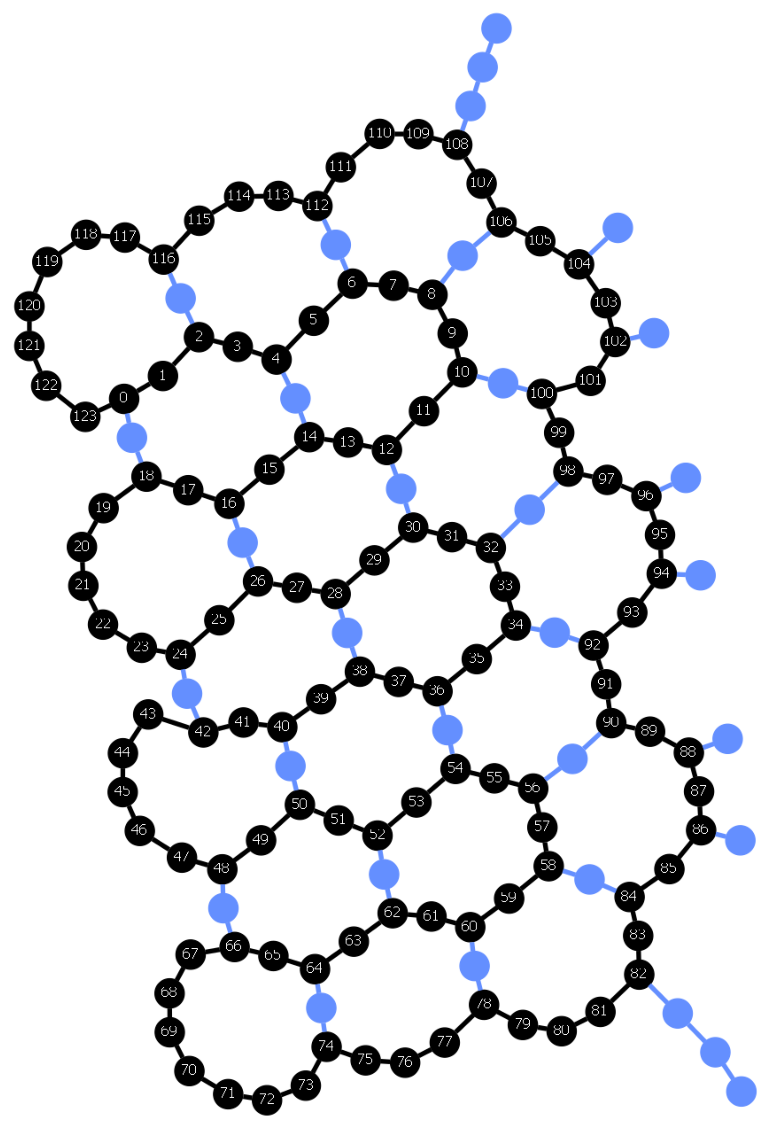}}
\caption{Map of all 156 qubits and their connectivity on \emph{ibm\_aachen}. The black nodes are the 124 qubits used in the experiments.}
\label{app:fig_aachen_layout}
\end{figure}

The circuit is a ring of 124 qubits as shown in \cref{app:fig_aachen_layout}. The unitary $U$ consists of a layer of Rot gates, followed by a ring of CNOTs and another layer of Rot gates, i.e. $U = U_{\text{ROT}} U_\text{ENT}  U_{\text{ROT}}$. Compared to the simulated experiments, the final entangling layers is omitted to reduce the influence of hardware noise.
This architecture leads to a total of 1488 rotational parameters, 744 weights and 744 bias parameters. Additionally, the model has 496 centroid parameters. However \ac{KAO} and \ac{CO} are not carried out due to the large runtime overhead required to estimate gradients using parameter shift rule.
The gate counts are given in \cref{tab:ibm_aachen_stats} and a vector graphic of the full circuit can be found in the supplementary material.
Since no training was carried out, the weights and bias were initialized with ones.

The model is compared to a SVC and an RBF kernel (named SimpleKernel) which is the classical equivalent of the untrained \textsc{McQuack} model.
The details of the SVC can be found in \cref{app:SVC_hyperparams} and the details of the RBF kernel in \cref{app:rbf_kernel_hyperparameters} 
The dataset used in the hardware experiments is described in \cref{app:hardware_dataset}.
Three independent runs were performed on each backend.

\begin{table}[htb]
\centering
\caption{Characterization of the ibm device used for the hardware experiments.}
\label{tab:ibm_aachen_stats}
\begin{tabular}{llcc}
\toprule
Category & Property & ibm\_pittsburgh & ibm\_aachen \\
\midrule
\multirow{5}{*}{System} 
  & Number of Qubits   & \multicolumn{2}{c}{156} \\
  & Processor Type     & Heron r3       & Heron r2 \\
  & QPU Version        & 1.0.8          & 1.0.0 \\
  & CLOPS              & \multicolumn{2}{c}{250{,}000} \\
  & Base Gates         & \multicolumn{2}{c}{CZ, ID, RX, RZ, RZZ, SX, X} \\
\midrule
\multirow{7}{*}{\shortstack{Error\\Metrics}}
  & Best 2Q Error      & $6.47 \times 10^{-4}$ & $8.57 \times 10^{-4}$ \\
  & 2Q Error (layered) & $2.79 \times 10^{-3}$ & $3.55 \times 10^{-3}$ \\
  & Median CZ Error    & $1.472 \times 10^{-3}$ & $1.848 \times 10^{-3}$ \\
  & Median SX Error    & $1.777 \times 10^{-4}$ & $1.88 \times 10^{-4}$ \\
  & Median Readout Error & $4.517 \times 10^{-3}$ & $7.568 \times 10^{-3}$ \\
  & Median $T_{1}$     & 301.81 $\mu$s  & 219.97 $\mu$s \\
  & Median $T_{2}$     & 339.51 $\mu$s  & 195.39 $\mu$s \\
\midrule
\multirow{5}{*}{\shortstack{Experimental\\Settings}}
  & Shots              & \multicolumn{2}{c}{100} \\
  & Error Mitigation   & \multicolumn{2}{c}{Dynamical Decoupling} \\
  &                    & \multicolumn{2}{c}{Measurement Twirling} \\
  & Bitflip Tolerance  & \multicolumn{2}{c}{45} \\
  & Optimization Level & \multicolumn{2}{c}{2} \\
\midrule
\multirow{4}{*}{\shortstack{Gate\\Counts}}
  & RZ                 & \multicolumn{2}{c}{1{,}612} \\
  & SX                 & \multicolumn{2}{c}{992} \\
  & CZ                 & \multicolumn{2}{c}{248} \\
  & Measure            & \multicolumn{2}{c}{124} \\
\bottomrule
\end{tabular}

\end{table}

\subsubsection{Bit Flip Tolerance Calibration} \label{app:bitflip_tolerance_calibration}

Bit flip tolerance \cite{bitflip_tolerance} is a pragmatic error mitigation technique for fidelity kernels.  
The entries of such fidelity kernels correspond to fidelities between pairs of data points and are estimated from the probability of sampling the all-zero bitstring $0^n$ from the associated quantum circuit, as defined in \cref{eq:quantum_kernel_hardware}. 
However, as the number of qubits and the overall noise level increase, this probability rapidly decreases, even when estimating the fidelity of a data point with itself. 

To address this issue, bit flip tolerance introduces a threshold parameter $b$. 
All measured bitstrings $s$ with Hamming weight $||s||_H \leq b$ are counted as contributions to the $0^n$ outcome. 
In this way, small deviations from the ideal all-zero measurement are attributed to noise rather than genuine quantum effects, thereby stabilizing the fidelity estimate.
The bit flip tolerance modifies the kernel expression of \cref{eq:quantum_kernel_hardware} to
\begin{align*}
k^{b}(\boldsymbol{x}_i,\boldsymbol{x}_j) 
&= \sum_{\|s\|_{H} \leq b} 
    \left| \langle s | U^\dagger (\boldsymbol{x}_j) U(\boldsymbol{x}_i)| 0^{n} \rangle \right|^{2}.
\end{align*}

\begin{figure}[H]
\centerline{\includegraphics[width=0.5\textwidth]{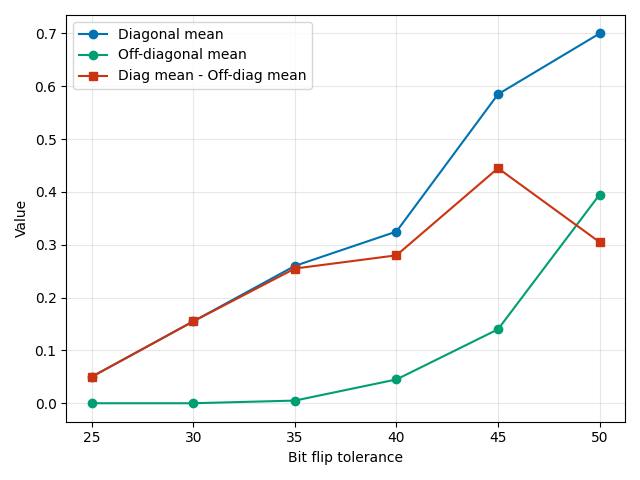}}
\caption{Bit flip tolerance calibration on ibm\_pittsburgh based on kernel estimates between class centroids for varying thresholds $b$. For a noise-free kernel, the diagonal mean (blue) equals one and the off-diagonal mean (green) is close to zero. The red curve depicts their difference, which serves as the calibration criterion $\overline{K^b_{\mathrm{diag}}} - \overline{K^b_{\mathrm{off}}}$. The maximum occurs at $b=45$, which is therefore chosen as the threshold in the hardware experiments.}
\label{app:fig_bitflip_calibration}
\end{figure}

We calibrate the bit-flip tolerance parameter $b$ by first initializing the centroids as the class means of the training data. Using these centroids, we estimate a kernel $K \in \mathbb{R}^{2 \times 2}$. This procedure is repeated for different values of $b$, and for each case we compute the mean of the diagonal $\overline{K^b_{\mathrm{diag}}}$ and off-diagonal entries $\overline{K^b_{\mathrm{off}}}$. These quantities, together with their difference, are shown in \cref{app:fig_bitflip_calibration}. In the noise-free setting, the diagonal elements equal one while the off-diagonal elements are substantially smaller, resulting in a large difference between the two. Accordingly, we select the value of $b$ that maximizes the difference between diagonal and off-diagonal entries of the kernel:
\begin{align*}
    b^{*} = \argmax_{b} \left( \overline{K^b_{\mathrm{diag}}} - \overline{K^b_{\mathrm{off}}} \right).
\end{align*}
As we can see in \cref{app:fig_bitflip_calibration}, the optimal choice is $b = 45$.

\subsubsection{Manifold Dataset} \label{app:hardware_dataset}
Benchmarking ideas in \ac{QML} is challenging since results are strongly influenced by experimental design choices, the limited system sizes currently accessible, and narratives shaped by the commercialization of quantum technologies \cite{schuld_qml_benchmarks2024}. 
To mitigate these issues, we adopt one of the data generation procedures proposed in the latter reference. 

Our chosen dataset is motivated by the manifold hypothesis \cite{manifold_1, manifold_2, manifold_3}, which suggests that data in modern machine learning predominantly lies on low-dimensional manifolds. 
Concretely, we generate inputs on a low-dimensional manifold and label them by a simple neural network initialized at random.
The inputs are then projected into a higher-dimensional space. 
The dataset parameters are summarized in \cref{tab:hardware_ds_params}.

\begin{table*}[htbp]
\centering
\caption{Parameters of the hidden manifold dataset used for the hardware experiments.}\label{tab:hardware_ds_params}
\begin{tabular}{lr}
\toprule
Parameter & Value  \\
\midrule
Manifold dimension & 5 \\
Feature dimension & 248 \\
Feature range & $[-1,1]$ \\
\# train samples & 100 \\
\# test samples & 50 \\
\bottomrule
\end{tabular}

\end{table*}

\subsubsection{SVC for Hardware Experiments} \label{app:SVC_hyperparams}
The SVC is one of the models the untrained McQuack is benchmarked against. 
The SVC used cross validation with 5 folds and 10 runs to find the best hyperparameters. The final hyperparameters are listed below.
\begin{verbatim}
    {'svc_C': 10, 'svc_gamma': 'scale', 'svc_kernel': 'rbf'}
\end{verbatim}

\subsubsection{RBF Kernel for Hardware Experiments} \label{app:rbf_kernel_hyperparameters}
The RBF kernel is a untrained classical baseline that calculates the RBF kernel between data and centroids. As for McQuack, the predicted label is the column number with the highest value for each sample.
This model has only one hyperparameter, the \texttt{scale} which relates to the \texttt{gamma} parameter of the SVC as $\gamma = \frac{1}{2 \,\text{scale}^2}$.
The RBF kernel is decaying with increasing distance between the points and this parameter only changes the width of the kernel. This means, it does not change the ordering of the kernel values. In other words, the index of the largest value in each row is independent from that parameter. Hence, no hyperparameter search was carried out and the parameter was arbitrarily set to $\texttt{scale} = 10$.

\subsection{Benchmarks}
The performance of \textsc{McQuack} is evaluated in two separate benchmarks. 
First, our model is evaluated on the QML Benchmark suite \cite{schuld_qml_benchmarks2024}, and second, it is tested on 19 real-world datasets.

\subsubsection{McQuack} \label{subsec:experiments_mcquack}
Due to the large number of datasets, our model is executed using a PyTorch-based, batch-wise state vector simulator that runs on a single NVIDIA DGX A100 with $\approx 80$ GB of RAM.
The centroids are initialized as the mean of the train samples for the respective class.
The embedding parameters $\boldsymbol{w}$ and $\boldsymbol{b}$ are initialized with variance 
$\gamma^2 = \frac{1}{3nL}$ (if hyperparameter \texttt{init\_small} equals True) for $L$ layers inspired by \cite{gaussian_init} which has shown this initialization has a lower bound of the expectation of the gradient norm of poly(n, L)$^{-1}$.
For comparison, the alternative initialization sets all weights and bias to one. 
The number of layers is controlled by the hyperparameter \texttt{n\_repetitions}, and setting it to one encodes each feature in (at least) one rotation gate.
The alternative initialization strategy (\texttt{init\_strategy}) sets the weights and bias such that for the first batch, the average angle is zero, i.e. $w \cdot x^{(d)} + b = 0$, where $x^{(d)}$ is the mean of the first batch of the $d$-th feature. This ensures that the angles in the rotation gates are close to zero at the initial optimization step. The motivation behind this initialization strategy is that it facilitates the first optimization step and ensures the parameters get an initial boost in the right direction.
A grid search was carried out to find the optimal hyperparameters of \textsc{McQuack} for the QML Benchmarks suite and the real world datasets. The respective search spaces are listed in \cref{app:tab:real_world_hyperparams,app:tab:qml_benchmarks_hyperparams} in \cref{sec:appC:benchmarks}.

\subsubsection{QML Benchmark Suite}

\cite{schuld_qml_benchmarks2024} points out the huge impact of the experimental design on the results and introduce a benchmarking suite consisting of 160 individual datasets from 6 binary classification tasks to better judge the ideas in \ac{QML}. Where not explicitly stated, our benchmarks follows their protocol.
We compare our model to the Quantum Neural Networks model family which is evaluated on 133 individual datasets from 5 classification tasks: linearly separable, downscaled MNIST, hidden manifolds, two curves, and hyperplanes.
The hyperparameter search space is listed in \cref{app:tab:qml_benchmarks_hyperparams} in \cref{subsec:appC:McQuack_hyperparameters}.
To speed up the hyperparameter search for the downscaled MNIST, a subset of 500 samples for training and 250 for testing were randomly selected for each fold of the cross validation.
Model rankings are computed from accuracies averaged over 5 random seeds after carrying out a hyperparameter search.
The other models in Quantum Neural Networks model family are:
\textsc{DressedQuantumCircuitClassifier} \cite{DressedQuantumCircuitClassifier},
\textsc{QuantumBoltzmannMachine} \cite{QuantumBoltzmannMachine},
\textsc{DataReuploadingClassifier} \cite{reupload_encoding_2020},
\textsc{IQPVariationalClassifier} \cite{Havl_ek_2019},
\textsc{QuantumMetricLearner} \cite{QuantumMetricLearner},
\textsc{CircuitCentricClassifier} \cite{CircuitCentricClassifier}, and
\textsc{TreeTensorClassifier} \cite{TreeTensorClassifier}.
A more detailed description of these models can be found in \cite{schuld_qml_benchmarks2024}.
The classical baseline in \cite{schuld_qml_benchmarks2024} is a multi-layer perceptron with two hidden layers of size 100.
Moreover, we employ \textsc{NystroemSVM}, a linear SVM with RBF kernel approximated by the Nystroem method.

\subsubsection{Real-World Datasets}
In addition, \textsc{McQuack} is evaluated on 19 real-world datasets from two to ten classes covering a wide range of applications which are described in \cref{app:tab:datasets_overview_description,app:tab:datasets_overview_ratios} in \cref{app:real_world_datasets}. Additionally, the manifold dataset from the hardware experiments is included in the benchmark as well.
The training set size is 1000 and the test set size 400. 
The hyperparameters of \textsc{McQuack} on the real-world datasets are listed in \cref{app:tab:real_world_hyperparams} in \cref{subsec:appC:McQuack_hyperparameters}.
Model rankings are computed from accuracies averaged over 5 random seeds.
The following models were chosen as baseline and their hyperparameters are described in \cref{app:hyperparameters}: 
\begin{enumerate}
    \item \acp{VC}, an ensemble of \acp{VC} using the same unitary as McQuack. The model is described in \cref{subsec:map_to_vc} and does not use shared parameters. The base classifier is based on \cite{wendlinger_adv_robustness}.
    \item SVC, a Support Vector Classifier based on sklearn's SVC\footnote{\url{https://scikit-learn.org/stable/modules/generated/sklearn.svm.SVC.html}}.
    \item \textsc{NystroemSVM}, a linear \ac{SVC} that uses the Nystroem method to approximate an RBF kernel. 
    \item NN, a two layer feedforward neural network, where the numbers of neurons is selected such that the total number of parameters is the same or slightly larger than that of McQuack. 
\end{enumerate}

\clearpage
\section{Benchmarks} \label{sec:appC:benchmarks}

\subsection{QML Benchmark Suite Additional Results} \label{subsec:appC:qml_benchmark_suite}

\begin{figure}[ht]
    \centering
    \includegraphics[width=0.45\textwidth]{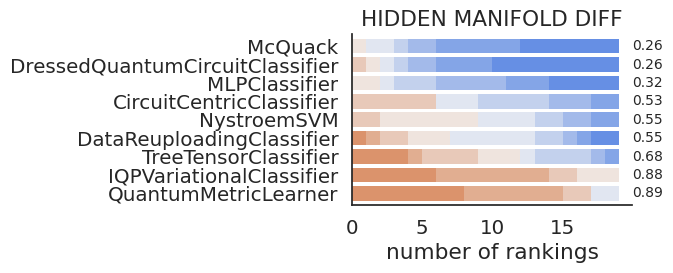}
    \includegraphics[width=0.45\textwidth]{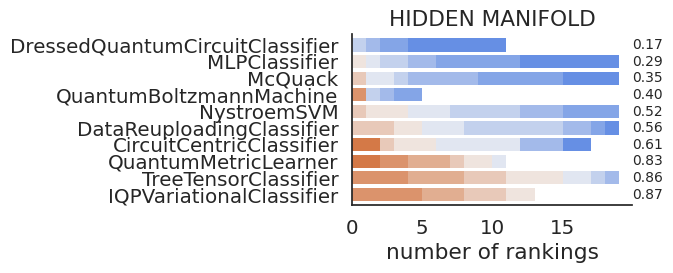}
    \includegraphics[width=0.45\textwidth]{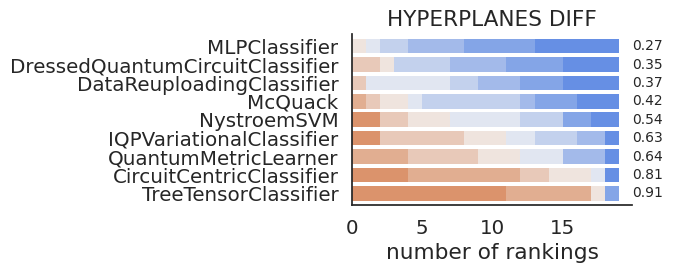}
    \includegraphics[width=0.45\textwidth]{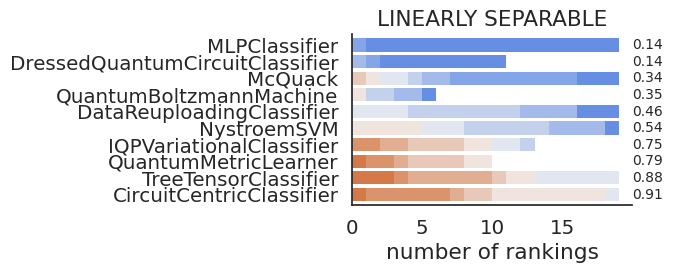}
    \includegraphics[width=0.45\textwidth]{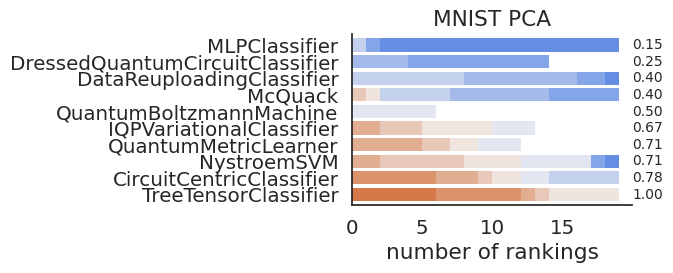}
    \includegraphics[width=0.45\textwidth]{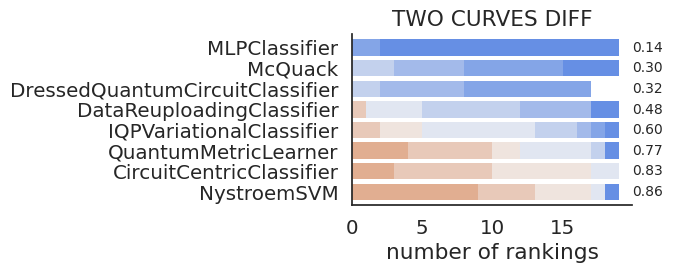}
    \includegraphics[width=0.45\textwidth]{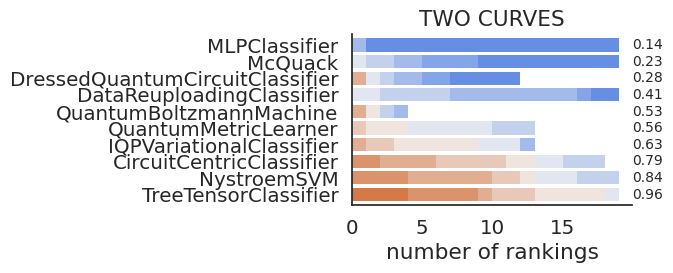}
    \caption{Rankings for the Quantum Neural Network family split up by classification tasks. 
    These rankings are largely consistent across the tasks. In 5 out of 7 benchmarks, the top three ranks are made up of the \textsc{MLPClassifier}, \textsc{DressedQuantumCircuitClassifier}, and \textsc{McQuack} in varying order. In \textsc{hyperplanes diff}, our model is put on fourth place by the \textsc{DataReuploadingClassifier} and in \textsc{linearly separable} by the \textsc{QuantumBoltzmannMachine}. However, the latter has been evaluated only on roughly a third of the datasets in that benchmark.
    }
    \label{app:fig:rankings-benchmarks_datasets}
\end{figure}

\begin{figure}[ht]
\centering
\includegraphics[width=0.45\textwidth]{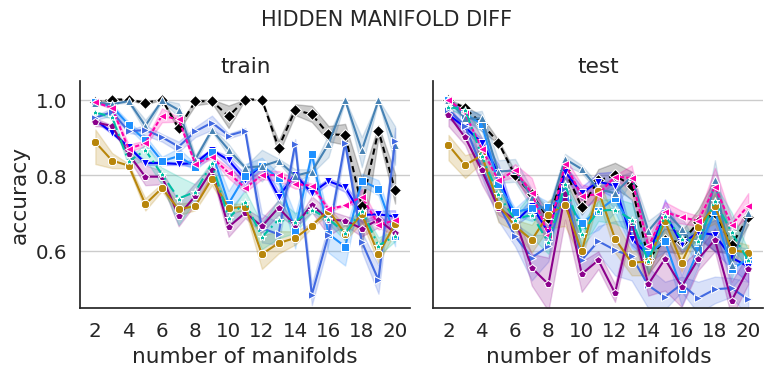}
\includegraphics[width=0.45\textwidth]{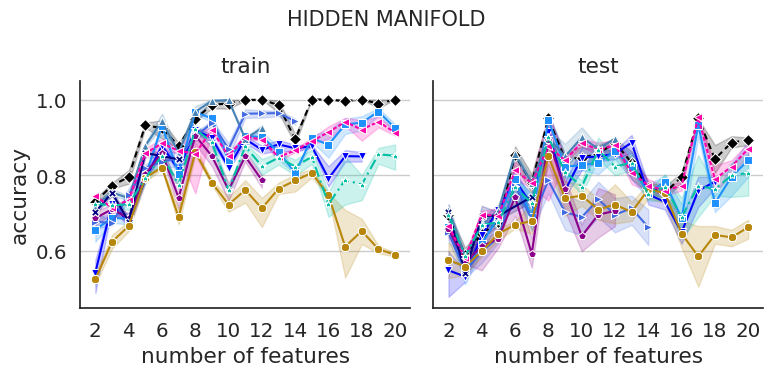}
\includegraphics[width=0.45\textwidth]{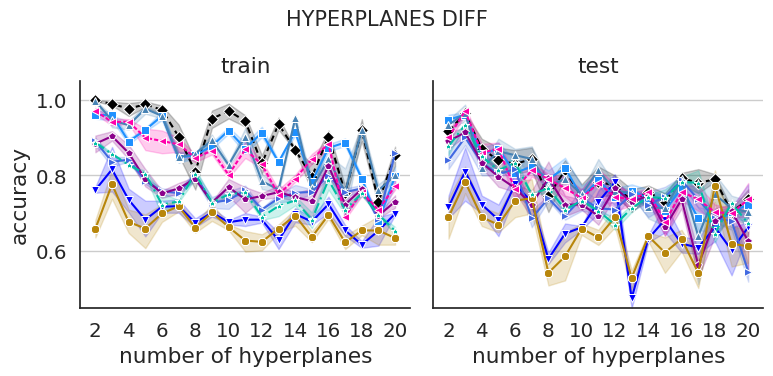}
\includegraphics[width=0.45\textwidth]{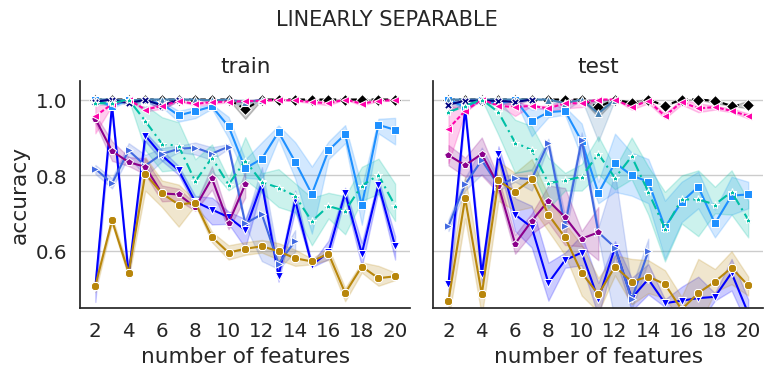}
\includegraphics[width=0.45\textwidth]{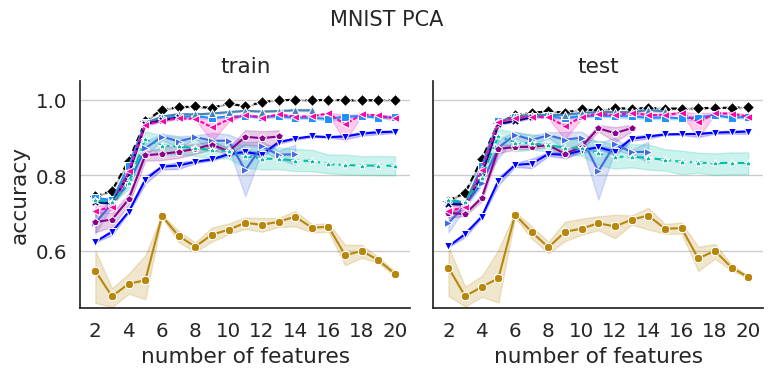}
\includegraphics[width=0.45\textwidth]{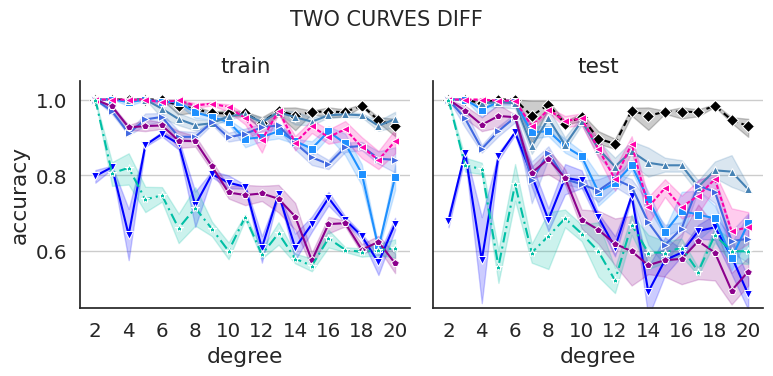}
\includegraphics[width=0.45\textwidth]{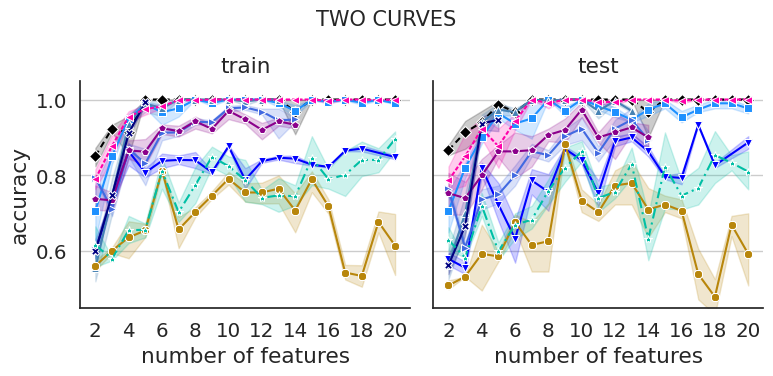}
\includegraphics[width=1.0\textwidth]{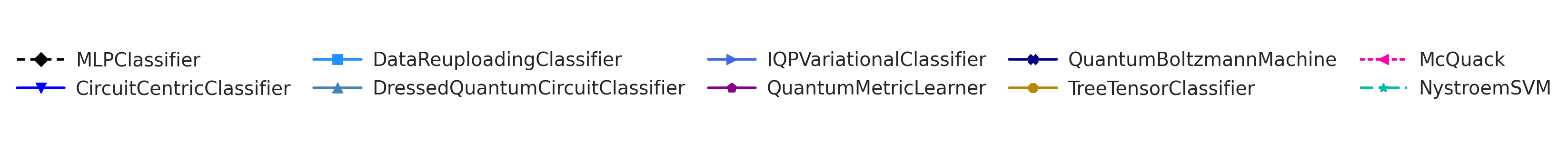}
\caption{Accuracy scores on the training and test sets are shown as a function of dataset dimensionality. The shading represents the 95\% confidence interval.
The classical \textsc{MLPClassifier} usually performs at the same level as or better than the quantum models.
\textsc{McQuack}, \textsc{DressedQuantumCircuitClassifier} and to a certain extend, \textsc{DataReuploadingClassifier} achieve high accuracy on most datasets while the remaining models perform considerably worse. 
Overfitting can be observed for \textsc{McQuack} on \textsc{hidden manifold} and \textsc{two curves diff}. 
However, the \textsc{MLPClassifier} and \textsc{DressedQuantumCircuitClassifier} struggle more with this phenomenon, achieving substantially lower test accuracy relative to training accuracy on \textsc{hidden manifold diff}, \textsc{hidden manifold} and \textsc{hyperplanes diff}. The latter model also overfits on \textsc{two curves diff}. 
Additionally, \textsc{McQuack} achieves test accuracies that are close to the highest among all models across most datasets, with \textsc{two curves diff} being the main exception. Therefore, we conclude that \textsc{McQuack} is less prone to overfitting than the other models, including the \textsc{MLPClassifier}.
\textsc{NystroemSVM} performs in the middle range and exhibits high variance across all datasets. This high variance is expected since the model subsamples only six landmark points for the kernel approximation.
There is no clear dependency of our model's performance on the dataset's dimensionality. On \textsc{hidden manifold diff}, \textsc{hyperplanes diff} and \textsc{two curves diff}, the performance decreases with an increasing number of dimensions, following the trend of the other quantum and classical models. However, on \textsc{hidden manifold}, \textsc{mnist pca} and \textsc{two curves}, performance increases with an increasing number of dimensions, again following the trend of the other models. Finally, on \textsc{linearly separable}, the performance of our model, \textsc{DressedQuantumCircuitClassifier}, \textsc{QuantumBoltzmannMachine}, and the classical model is good and largely stable, whereas all other quantum models struggle. 
The plot is adapted from \cite{schuld_qml_benchmarks2024}.
}
\label{app:fig:scores-datasets}
\end{figure}

\clearpage 

\subsection{Real-world Datasets} \label{app:real_world_datasets}

\begin{table*}[h]
\centering
\caption{Overview of the real-world datasets used in the benchmark.}\label{app:tab:datasets_overview_description}
\begingroup
\setlength{\tabcolsep}{4pt}
\footnotesize
\begin{tabularx}{\linewidth}{lllr>{\raggedright\arraybackslash}X}
\toprule
Dataset & Ref & Description & \# Features & Classes \\
\midrule
Census & \cite{census_income} & Income & 14 & $\leq50K,\allowbreak >50K$ \\
CoverT & \cite{cover_type} & Forest tree types & 15 & $\leq4,\allowbreak >4$ \\
DoH & \cite{DoH} & Network traffic & 33 & benign,\allowbreak malicious \\
EMNIST & \cite{emnist} & Handwritten letters & 784 & A-M,\allowbreak N-Z \\
EMNIST10 &  &  &  & A-J \\
EMNIST4 &  &  &  & A,B,C,D \\
FMNIST & \cite{fmnist} & Clothing types & 784 & 0-4,\allowbreak 5-9 \\
FMNIST10 &  &  &  & 0-9 \\
FMNIST5 &  &  &  & 0,1,2,3,4 \\
Iris3 & \cite{iris} & Iris flower species & 4 & setosa,\allowbreak versicolour,\allowbreak virginica \\
KDD & \cite{KDD} & Network intrusion & 42 & normal,\allowbreak anomalous \\
KDD10 &  &  &  & normal,\allowbreak neptune,\allowbreak satan,\allowbreak ipsweep,\allowbreak smurf,\allowbreak portsweep,\allowbreak nmap,\allowbreak back,\allowbreak guess\_passwd,\allowbreak mscan \\
KDD4 &  &  &  & normal,\allowbreak neptune,\allowbreak satan,\allowbreak ipsweep \\
MNIST & \cite{mnist} & Handwritten digits & 784 & 0-4,\allowbreak 5-9 \\
MNIST10 &  &  &  & 0-9 \\
MNIST4 &  &  &  & 0,1,2,3 \\
Manifold & \cite{schuld_qml_benchmarks2024} & Hidden manifold & 248 & 0,1 \\
PM2 & \cite{wendlinger_adv_robustness} & Images of plus/minus signs & 256 & $-,\allowbreak +$ \\
PM4 &  &  &  & $-,\allowbreak +,\allowbreak \vdash,\allowbreak \dashv$ \\
URL & \cite{URL} & URLs & 79 & benign,\allowbreak non-benign \\
\bottomrule
\end{tabularx}
\endgroup
\end{table*}

\begin{table*}[ht]
\centering
\caption{Class distributions of all benchmark datasets. Columns 0--9 denote the number of samples per class, and \texttt{Ratio} gives the ratio of the smallest class to the total number of samples in each split. Note that due to the small number of samples in \texttt{Iris3}, the validation and test sets are identical.}
\label{app:tab:datasets_overview_ratios}
\footnotesize
\begin{tabular}{llrrrrrrrrrrr}
\toprule
Dataset &  Split & Ratio &0 & 1 & 2 & 3 & 4 & 5 & 6 & 7 & 8 & 9 \\
\midrule
Census & train & 0.275 & 725 & 275 & 0 & 0 & 0 & 0 & 0 & 0 & 0 & 0 \\
  & val & 0.260 & 296 & 104 & 0 & 0 & 0 & 0 & 0 & 0 & 0 & 0 \\
  & test & 0.250 & 300 & 100 & 0 & 0 & 0 & 0 & 0 & 0 & 0 & 0 \\
CoverT & train & 0.473 & 473 & 527 & 0 & 0 & 0 & 0 & 0 & 0 & 0 & 0 \\
  & val & 0.477 & 191 & 209 & 0 & 0 & 0 & 0 & 0 & 0 & 0 & 0 \\
  & test & 0.500 & 200 & 200 & 0 & 0 & 0 & 0 & 0 & 0 & 0 & 0 \\
DoH & train & 0.241 & 759 & 241 & 0 & 0 & 0 & 0 & 0 & 0 & 0 & 0 \\
  & val & 0.228 & 309 & 91 & 0 & 0 & 0 & 0 & 0 & 0 & 0 & 0 \\
  & test & 0.265 & 294 & 106 & 0 & 0 & 0 & 0 & 0 & 0 & 0 & 0 \\
EMNIST & train & 0.496 & 496 & 504 & 0 & 0 & 0 & 0 & 0 & 0 & 0 & 0 \\
  & val & 0.480 & 208 & 192 & 0 & 0 & 0 & 0 & 0 & 0 & 0 & 0 \\
  & test & 0.492 & 203 & 197 & 0 & 0 & 0 & 0 & 0 & 0 & 0 & 0 \\
EMNIST10 & train & 0.078 & 108 & 89 & 85 & 108 & 89 & 109 & 106 & 78 & 109 & 119 \\
  & val & 0.075 & 46 & 47 & 35 & 38 & 40 & 30 & 48 & 35 & 44 & 37 \\
  & test & 0.075 & 40 & 34 & 39 & 40 & 30 & 42 & 43 & 42 & 39 & 51 \\
EMNIST4 & train & 0.234 & 268 & 235 & 263 & 234 & 0 & 0 & 0 & 0 & 0 & 0 \\
  & val & 0.228 & 91 & 107 & 91 & 111 & 0 & 0 & 0 & 0 & 0 & 0 \\
  & test & 0.225 & 90 & 111 & 107 & 92 & 0 & 0 & 0 & 0 & 0 & 0 \\
FMNIST & train & 0.499 & 501 & 499 & 0 & 0 & 0 & 0 & 0 & 0 & 0 & 0 \\
  & val & 0.487 & 195 & 205 & 0 & 0 & 0 & 0 & 0 & 0 & 0 & 0 \\
  & test & 0.472 & 189 & 211 & 0 & 0 & 0 & 0 & 0 & 0 & 0 & 0 \\
FMNIST10 & train & 0.085 & 108 & 85 & 105 & 99 & 102 & 94 & 105 & 106 & 92 & 104 \\
  & val & 0.072 & 32 & 46 & 37 & 40 & 50 & 45 & 46 & 41 & 34 & 29 \\
  & test & 0.075 & 46 & 38 & 41 & 53 & 33 & 34 & 36 & 30 & 49 & 40 \\
FMNIST5 & train & 0.174 & 174 & 219 & 216 & 200 & 191 & 0 & 0 & 0 & 0 & 0 \\
  & val & 0.182 & 80 & 73 & 87 & 75 & 85 & 0 & 0 & 0 & 0 & 0 \\
  & test & 0.177 & 71 & 90 & 86 & 71 & 82 & 0 & 0 & 0 & 0 & 0 \\
Iris3 & train & 0.295 & 31 & 37 & 37 & 0 & 0 & 0 & 0 & 0 & 0 & 0 \\
  & val & 0.289 & 19 & 13 & 13 & 0 & 0 & 0 & 0 & 0 & 0 & 0 \\
  & test & 0.289 & 19 & 13 & 13 & 0 & 0 & 0 & 0 & 0 & 0 & 0 \\
KDD & train & 0.496 & 504 & 496 & 0 & 0 & 0 & 0 & 0 & 0 & 0 & 0 \\
  & val & 0.443 & 223 & 177 & 0 & 0 & 0 & 0 & 0 & 0 & 0 & 0 \\
  & test & 0.487 & 205 & 195 & 0 & 0 & 0 & 0 & 0 & 0 & 0 & 0 \\
KDD10 & train & 0.089 & 96 & 110 & 109 & 97 & 107 & 103 & 95 & 94 & 89 & 100 \\
  & val & 0.075 & 44 & 44 & 47 & 33 & 42 & 35 & 48 & 30 & 32 & 45 \\
  & test & 0.080 & 53 & 40 & 36 & 43 & 34 & 47 & 39 & 32 & 42 & 34 \\
KDD4 & train & 0.239 & 239 & 242 & 264 & 255 & 0 & 0 & 0 & 0 & 0 & 0 \\
  & val & 0.225 & 113 & 90 & 100 & 97 & 0 & 0 & 0 & 0 & 0 & 0 \\
  & test & 0.242 & 102 & 100 & 97 & 101 & 0 & 0 & 0 & 0 & 0 & 0 \\
MNIST & train & 0.482 & 482 & 518 & 0 & 0 & 0 & 0 & 0 & 0 & 0 & 0 \\
  & val & 0.468 & 213 & 187 & 0 & 0 & 0 & 0 & 0 & 0 & 0 & 0 \\
  & test & 0.465 & 186 & 214 & 0 & 0 & 0 & 0 & 0 & 0 & 0 & 0 \\
MNIST10 & train & 0.080 & 108 & 129 & 96 & 117 & 82 & 89 & 92 & 100 & 107 & 80 \\
  & val & 0.083 & 38 & 37 & 36 & 40 & 40 & 41 & 42 & 49 & 33 & 44 \\
  & test & 0.090 & 36 & 39 & 41 & 41 & 42 & 45 & 36 & 42 & 39 & 39 \\
MNIST4 & train & 0.210 & 210 & 299 & 245 & 246 & 0 & 0 & 0 & 0 & 0 & 0 \\
  & val & 0.225 & 90 & 116 & 100 & 94 & 0 & 0 & 0 & 0 & 0 & 0 \\
  & test & 0.220 & 101 & 107 & 104 & 88 & 0 & 0 & 0 & 0 & 0 & 0 \\
Manifold & train & 0.471 & 471 & 529 & 0 & 0 & 0 & 0 & 0 & 0 & 0 & 0 \\
  & val & 0.463 & 185 & 215 & 0 & 0 & 0 & 0 & 0 & 0 & 0 & 0 \\
  & test & 0.485 & 206 & 194 & 0 & 0 & 0 & 0 & 0 & 0 & 0 & 0 \\
PM2 & train & 0.500 & 250 & 250 & 0 & 0 & 0 & 0 & 0 & 0 & 0 & 0 \\
  & val & 0.500 & 50 & 50 & 0 & 0 & 0 & 0 & 0 & 0 & 0 & 0 \\
  & test & 0.500 & 50 & 50 & 0 & 0 & 0 & 0 & 0 & 0 & 0 & 0 \\
PM4 & train & 0.250 & 250 & 250 & 250 & 250 & 0 & 0 & 0 & 0 & 0 & 0 \\
  & val & 0.250 & 50 & 50 & 50 & 50 & 0 & 0 & 0 & 0 & 0 & 0 \\
  & test & 0.250 & 50 & 50 & 50 & 50 & 0 & 0 & 0 & 0 & 0 & 0 \\
URL & train & 0.496 & 496 & 504 & 0 & 0 & 0 & 0 & 0 & 0 & 0 & 0 \\
  & val & 0.470 & 212 & 188 & 0 & 0 & 0 & 0 & 0 & 0 & 0 & 0 \\
  & test & 0.475 & 210 & 190 & 0 & 0 & 0 & 0 & 0 & 0 & 0 & 0 \\
\hline
\end{tabular}
\end{table*}

\begingroup
\setlength{\tabcolsep}{2pt}
\begin{table*}[t]
\centering
\caption{Results across datasets. Bold values indicate the highest mean performance for the metric while underlined values denote the second-highest.}\label{tab:results_real_world_ds}
\footnotesize
\begin{tabular}{llrrrrrrr}
\toprule
Dataset & Model & Params. & AUC & Precision & Recall & F1 & Acc. & Bal.Acc. \\
\midrule
Census & \underline{McQuack} & 136 & $\underline{0.87 \pm 0.01}$ & $\underline{0.75 \pm 0.07}$ & $0.54 \pm 0.03$ & $0.51 \pm 0.06$ & $0.77 \pm 0.01$ & $0.54 \pm 0.03$ \\
 & VC & 216 & $0.52 \pm 0.02$ & $0.38 \pm 0.00$ & $0.50 \pm 0.00$ & $0.43 \pm 0.00$ & $0.75 \pm 0.00$ & $0.50 \pm 0.00$ \\
 & \textbf{SVC} & 558 & $\underline{0.87 \pm 0.00}$ & $\bm{0.76 \pm 0.02}$ & $\underline{0.72 \pm 0.01}$ & $\underline{0.73 \pm 0.01}$ & $\underline{0.81 \pm 0.01}$ & $\underline{0.72 \pm 0.01}$ \\
 & \textbf{NN} & 137 & $\bm{0.89 \pm 0.01}$ & $\bm{0.76 \pm 0.01}$ & $\bm{0.74 \pm 0.02}$ & $\bm{0.75 \pm 0.02}$ & $\bm{0.82 \pm 0.01}$ & $\bm{0.74 \pm 0.02}$ \\
 & NystroemSVM & 7 & $0.80 \pm 0.04$ & $0.73 \pm 0.06$ & $0.61 \pm 0.08$ & $0.61 \pm 0.11$ & $0.78 \pm 0.03$ & $0.61 \pm 0.08$ \\
\midrule
CoverT & \underline{McQuack} & 138 & $\underline{0.78 \pm 0.03}$ & $\underline{0.70 \pm 0.03}$ & $\underline{0.69 \pm 0.03}$ & $\underline{0.68 \pm 0.03}$ & $\underline{0.69 \pm 0.03}$ & $\underline{0.69 \pm 0.03}$ \\
 & VC & 216 & $0.57 \pm 0.00$ & $0.25 \pm 0.00$ & $0.50 \pm 0.00$ & $0.33 \pm 0.00$ & $0.50 \pm 0.00$ & $0.50 \pm 0.00$ \\
 & SVC & 832 & $0.68 \pm 0.00$ & $0.66 \pm 0.00$ & $0.64 \pm 0.00$ & $0.63 \pm 0.00$ & $0.64 \pm 0.00$ & $0.64 \pm 0.00$ \\
 & \textbf{NN} & 143 & $\bm{0.80 \pm 0.05}$ & $\bm{0.76 \pm 0.02}$ & $\bm{0.74 \pm 0.04}$ & $\bm{0.74 \pm 0.04}$ & $\bm{0.74 \pm 0.04}$ & $\bm{0.74 \pm 0.04}$ \\
 & NystroemSVM & 7 & $0.71 \pm 0.02$ & $0.66 \pm 0.02$ & $0.64 \pm 0.03$ & $0.63 \pm 0.04$ & $0.64 \pm 0.03$ & $0.64 \pm 0.03$ \\
\midrule
DoH & \underline{McQuack} & 282 & $\underline{0.94 \pm 0.01}$ & $\underline{0.88 \pm 0.01}$ & $0.83 \pm 0.01$ & $0.85 \pm 0.01$ & $0.89 \pm 0.01$ & $0.83 \pm 0.01$ \\
 & VC & 432 & $0.59 \pm 0.02$ & $0.37 \pm 0.00$ & $0.50 \pm 0.00$ & $0.42 \pm 0.00$ & $0.73 \pm 0.00$ & $0.50 \pm 0.00$ \\
 & SVC & 483 & $0.92 \pm 0.00$ & $0.83 \pm 0.00$ & $0.83 \pm 0.00$ & $0.83 \pm 0.00$ & $0.87 \pm 0.00$ & $0.83 \pm 0.00$ \\
 & \textbf{NN} & 290 & $\bm{0.97 \pm 0.00}$ & $\bm{0.94 \pm 0.01}$ & $\bm{0.90 \pm 0.01}$ & $\bm{0.92 \pm 0.01}$ & $\bm{0.94 \pm 0.01}$ & $\bm{0.90 \pm 0.01}$ \\
 & \underline{NystroemSVM} & 7 & $0.93 \pm 0.01$ & $\underline{0.88 \pm 0.01}$ & $\underline{0.84 \pm 0.01}$ & $\underline{0.86 \pm 0.00}$ & $\underline{0.90 \pm 0.00}$ & $\underline{0.84 \pm 0.01}$ \\
\midrule
EMNIST & McQuack & 6,284 & $0.76 \pm 0.01$ & $0.70 \pm 0.01$ & $0.70 \pm 0.01$ & $0.70 \pm 0.01$ & $0.70 \pm 0.01$ & $0.70 \pm 0.01$ \\
 & VC & 9,432 & $0.73 \pm 0.00$ & $0.69 \pm 0.01$ & $0.69 \pm 0.01$ & $0.69 \pm 0.01$ & $0.69 \pm 0.01$ & $0.69 \pm 0.01$ \\
 & \textbf{SVC} & 752 & $\bm{0.84 \pm 0.00}$ & $\bm{0.77 \pm 0.00}$ & $\bm{0.77 \pm 0.00}$ & $\bm{0.77 \pm 0.00}$ & $\bm{0.77 \pm 0.00}$ & $\bm{0.77 \pm 0.00}$ \\
 & \underline{NN} & 6,304 & $\underline{0.77 \pm 0.01}$ & $\underline{0.72 \pm 0.02}$ & $\underline{0.72 \pm 0.02}$ & $\underline{0.72 \pm 0.02}$ & $\underline{0.72 \pm 0.02}$ & $\underline{0.72 \pm 0.02}$ \\
 & NystroemSVM & 7 & $0.61 \pm 0.04$ & $0.58 \pm 0.04$ & $0.58 \pm 0.04$ & $0.57 \pm 0.05$ & $0.57 \pm 0.04$ & $0.58 \pm 0.04$ \\
\midrule
EMNIST10 & McQuack & 12,556 & $0.90 \pm 0.01$ & $0.61 \pm 0.04$ & $0.61 \pm 0.04$ & $0.60 \pm 0.04$ & $0.61 \pm 0.04$ & $0.61 \pm 0.04$ \\
 & \underline{VC} & 47,160 & $\underline{0.94 \pm 0.00}$ & $0.71 \pm 0.00$ & $\underline{0.71 \pm 0.00}$ & $\underline{0.70 \pm 0.00}$ & $\underline{0.71 \pm 0.00}$ & $\underline{0.71 \pm 0.00}$ \\
 & \textbf{SVC} & 8,721 & $\bm{0.96 \pm 0.00}$ & $\bm{0.87 \pm 0.01}$ & $0.65 \pm 0.02$ & $\underline{0.70 \pm 0.02}$ & $0.66 \pm 0.02$ & $0.65 \pm 0.02$ \\
 & \textbf{NN} & 12,840 & $\underline{0.94 \pm 0.01}$ & $\underline{0.72 \pm 0.03}$ & $\bm{0.72 \pm 0.03}$ & $\bm{0.71 \pm 0.03}$ & $\bm{0.72 \pm 0.03}$ & $\bm{0.72 \pm 0.03}$ \\
 & \underline{NystroemSVM} & 310 & $\underline{0.94 \pm 0.00}$ & $0.68 \pm 0.02$ & $0.68 \pm 0.02$ & $0.68 \pm 0.02$ & $0.68 \pm 0.02$ & $0.68 \pm 0.02$ \\
\midrule
EMNIST4 & \textbf{McQuack} & 7,852 & $\underline{0.96 \pm 0.00}$ & $\underline{0.85 \pm 0.01}$ & $\bm{0.85 \pm 0.01}$ & $\bm{0.85 \pm 0.01}$ & $\bm{0.85 \pm 0.01}$ & $\bm{0.85 \pm 0.01}$ \\
 & \underline{VC} & 18,864 & $\underline{0.96 \pm 0.00}$ & $0.82 \pm 0.00$ & $0.82 \pm 0.00$ & $0.82 \pm 0.00$ & $0.82 \pm 0.00$ & $0.82 \pm 0.00$ \\
 & \textbf{SVC} & 3,000 & $\bm{0.98 \pm 0.00}$ & $\bm{0.86 \pm 0.01}$ & $0.81 \pm 0.02$ & $0.81 \pm 0.02$ & $0.81 \pm 0.02$ & $0.81 \pm 0.02$ \\
 & \underline{NN} & 7,914 & $\underline{0.96 \pm 0.00}$ & $0.84 \pm 0.01$ & $\underline{0.84 \pm 0.01}$ & $\underline{0.83 \pm 0.01}$ & $\underline{0.84 \pm 0.01}$ & $\underline{0.84 \pm 0.01}$ \\
 & NystroemSVM & 52 & $0.91 \pm 0.01$ & $0.72 \pm 0.04$ & $0.72 \pm 0.04$ & $0.71 \pm 0.04$ & $0.73 \pm 0.04$ & $0.72 \pm 0.04$ \\
\midrule
FMNIST & \textbf{McQuack} & 6,284 & $\bm{0.96 \pm 0.01}$ & $\underline{0.90 \pm 0.01}$ & $\underline{0.90 \pm 0.01}$ & $\underline{0.90 \pm 0.01}$ & $\underline{0.90 \pm 0.01}$ & $\underline{0.90 \pm 0.01}$ \\
 & \underline{VC} & 9,432 & $\underline{0.95 \pm 0.00}$ & $\underline{0.90 \pm 0.00}$ & $0.89 \pm 0.00$ & $0.89 \pm 0.00$ & $0.89 \pm 0.00$ & $0.89 \pm 0.00$ \\
 & \underline{SVC} & 999 & $\underline{0.95 \pm 0.00}$ & $0.79 \pm 0.01$ & $0.74 \pm 0.03$ & $0.72 \pm 0.04$ & $0.73 \pm 0.03$ & $0.74 \pm 0.03$ \\
 & \textbf{NN} & 6,304 & $\bm{0.96 \pm 0.00}$ & $\bm{0.91 \pm 0.00}$ & $\bm{0.91 \pm 0.00}$ & $\bm{0.91 \pm 0.00}$ & $\bm{0.91 \pm 0.00}$ & $\bm{0.91 \pm 0.00}$ \\
 & NystroemSVM & 7 & $0.93 \pm 0.01$ & $0.85 \pm 0.02$ & $0.85 \pm 0.02$ & $0.85 \pm 0.02$ & $0.85 \pm 0.02$ & $0.85 \pm 0.02$ \\
\midrule
FMNIST10 & \underline{McQuack} & 12,556 & $\underline{0.93 \pm 0.01}$ & $\underline{0.72 \pm 0.01}$ & $\underline{0.72 \pm 0.01}$ & $\underline{0.72 \pm 0.01}$ & $\underline{0.73 \pm 0.01}$ & $\underline{0.72 \pm 0.01}$ \\
 & \textbf{VC} & 47,160 & $\bm{0.95 \pm 0.00}$ & $0.69 \pm 0.01$ & $0.71 \pm 0.00$ & $0.69 \pm 0.01$ & $0.72 \pm 0.00$ & $0.71 \pm 0.00$ \\
 & SVC & 9,000 & $0.92 \pm 0.01$ & $0.62 \pm 0.05$ & $0.60 \pm 0.04$ & $0.56 \pm 0.04$ & $0.60 \pm 0.04$ & $0.60 \pm 0.04$ \\
 & \textbf{NN} & 12,840 & $\bm{0.95 \pm 0.01}$ & $\bm{0.77 \pm 0.01}$ & $\bm{0.77 \pm 0.01}$ & $\bm{0.76 \pm 0.01}$ & $\bm{0.77 \pm 0.01}$ & $\bm{0.77 \pm 0.01}$ \\
 & \textbf{NystroemSVM} & 310 & $\bm{0.95 \pm 0.00}$ & $\underline{0.72 \pm 0.02}$ & $\underline{0.72 \pm 0.02}$ & $0.71 \pm 0.02$ & $\underline{0.73 \pm 0.02}$ & $\underline{0.72 \pm 0.02}$ \\
\midrule
FMNIST5 & \underline{McQuack} & 8,636 & $\underline{0.94 \pm 0.01}$ & $0.77 \pm 0.01$ & $\underline{0.77 \pm 0.01}$ & $0.77 \pm 0.02$ & $\underline{0.77 \pm 0.02}$ & $\underline{0.77 \pm 0.01}$ \\
 & \textbf{VC} & 23,580 & $\bm{0.95 \pm 0.00}$ & $\underline{0.79 \pm 0.01}$ & $\underline{0.77 \pm 0.01}$ & $\underline{0.78 \pm 0.01}$ & $\underline{0.77 \pm 0.01}$ & $\underline{0.77 \pm 0.01}$ \\
 & SVC & 3,928 & $0.93 \pm 0.01$ & $0.73 \pm 0.01$ & $0.73 \pm 0.01$ & $0.72 \pm 0.03$ & $0.73 \pm 0.01$ & $0.73 \pm 0.01$ \\
 & \textbf{NN} & 8,725 & $\bm{0.95 \pm 0.01}$ & $\bm{0.81 \pm 0.01}$ & $\bm{0.80 \pm 0.01}$ & $\bm{0.80 \pm 0.01}$ & $\bm{0.80 \pm 0.01}$ & $\bm{0.80 \pm 0.01}$ \\
 & \textbf{NystroemSVM} & 80 & $\bm{0.95 \pm 0.00}$ & $0.76 \pm 0.01$ & $0.76 \pm 0.01$ & $0.76 \pm 0.01$ & $0.76 \pm 0.01$ & $0.76 \pm 0.01$ \\
\midrule
Iris3 & \underline{McQuack} & 48 & $\underline{0.98 \pm 0.00}$ & $0.89 \pm 0.01$ & $0.83 \pm 0.01$ & $0.81 \pm 0.01$ & $0.85 \pm 0.01$ & $0.83 \pm 0.01$ \\
 & VC & 108 & $0.83 \pm 0.02$ & $0.46 \pm 0.27$ & $0.50 \pm 0.19$ & $0.37 \pm 0.24$ & $0.49 \pm 0.23$ & $0.50 \pm 0.19$ \\
 & \textbf{SVC} & 48 & $\bm{1.00 \pm 0.00}$ & $\underline{0.99 \pm 0.01}$ & $\underline{0.98 \pm 0.01}$ & $\underline{0.98 \pm 0.01}$ & $\underline{0.99 \pm 0.01}$ & $\underline{0.98 \pm 0.01}$ \\
 & NN & 55 & $0.97 \pm 0.07$ & $0.72 \pm 0.21$ & $0.74 \pm 0.08$ & $0.68 \pm 0.13$ & $0.77 \pm 0.07$ & $0.74 \pm 0.08$ \\
 & \textbf{NystroemSVM} & 30 & $\bm{1.00 \pm 0.00}$ & $\bm{1.00 \pm 0.00}$ & $\bm{1.00 \pm 0.00}$ & $\bm{1.00 \pm 0.00}$ & $\bm{1.00 \pm 0.00}$ & $\bm{1.00 \pm 0.00}$ \\
\midrule
\multicolumn{9}{r}{\textit{Continued on next page}} \\
\bottomrule
\end{tabular}
\end{table*}
\endgroup

\begingroup
\setlength{\tabcolsep}{2pt}
\begin{table*}[t]
\centering
\footnotesize
\begin{tabular}{llrrrrrrr}
\toprule
Dataset & Model & Params. & AUC & Precision & Recall & F1 & Acc. & Bal.Acc. \\
\midrule
KDD & \textbf{McQuack} & 336 & $\bm{0.99 \pm 0.00}$ & $\underline{0.95 \pm 0.01}$ & $\underline{0.95 \pm 0.01}$ & $\underline{0.95 \pm 0.01}$ & $\underline{0.95 \pm 0.01}$ & $\underline{0.95 \pm 0.01}$ \\
 & VC & 504 & $0.95 \pm 0.01$ & $0.86 \pm 0.02$ & $0.83 \pm 0.01$ & $0.83 \pm 0.01$ & $0.84 \pm 0.01$ & $0.83 \pm 0.01$ \\
 & \textbf{SVC} & 335 & $\bm{0.99 \pm 0.00}$ & $\underline{0.95 \pm 0.00}$ & $\underline{0.95 \pm 0.00}$ & $\underline{0.95 \pm 0.00}$ & $\underline{0.95 \pm 0.00}$ & $\underline{0.95 \pm 0.00}$ \\
 & \textbf{NN} & 343 & $\bm{0.99 \pm 0.00}$ & $\bm{0.97 \pm 0.01}$ & $\bm{0.97 \pm 0.01}$ & $\bm{0.97 \pm 0.01}$ & $\bm{0.97 \pm 0.01}$ & $\bm{0.97 \pm 0.01}$ \\
 & \underline{NystroemSVM} & 7 & $\underline{0.97 \pm 0.00}$ & $0.89 \pm 0.01$ & $0.89 \pm 0.01$ & $0.89 \pm 0.01$ & $0.89 \pm 0.01$ & $0.89 \pm 0.01$ \\
\midrule
KDD10 & \underline{McQuack} & 718 & $\underline{0.99 \pm 0.00}$ & $0.92 \pm 0.02$ & $0.91 \pm 0.02$ & $0.91 \pm 0.02$ & $0.91 \pm 0.02$ & $0.91 \pm 0.02$ \\
 & VC & 2,880 & $0.88 \pm 0.03$ & $0.20 \pm 0.14$ & $0.22 \pm 0.11$ & $0.14 \pm 0.11$ & $0.21 \pm 0.10$ & $0.22 \pm 0.11$ \\
 & \textbf{SVC} & 2,178 & $\bm{1.00 \pm 0.00}$ & $\bm{0.99 \pm 0.00}$ & $\bm{0.99 \pm 0.00}$ & $\bm{0.99 \pm 0.00}$ & $\bm{0.99 \pm 0.00}$ & $\bm{0.99 \pm 0.00}$ \\
 & \textbf{NN} & 736 & $\bm{1.00 \pm 0.00}$ & $\bm{0.99 \pm 0.00}$ & $\bm{0.99 \pm 0.00}$ & $\bm{0.99 \pm 0.00}$ & $\bm{0.99 \pm 0.00}$ & $\bm{0.99 \pm 0.00}$ \\
 & \textbf{NystroemSVM} & 310 & $\bm{1.00 \pm 0.00}$ & $\underline{0.98 \pm 0.01}$ & $\underline{0.98 \pm 0.01}$ & $\underline{0.98 \pm 0.01}$ & $\underline{0.98 \pm 0.01}$ & $\underline{0.98 \pm 0.01}$ \\
\midrule
KDD4 & \textbf{McQuack} & 460 & $\bm{1.00 \pm 0.00}$ & $\underline{0.99 \pm 0.00}$ & $\underline{0.99 \pm 0.00}$ & $\underline{0.99 \pm 0.00}$ & $\underline{0.99 \pm 0.00}$ & $\underline{0.99 \pm 0.00}$ \\
 & \underline{VC} & 1,152 & $\underline{0.94 \pm 0.01}$ & $0.65 \pm 0.15$ & $0.66 \pm 0.07$ & $0.60 \pm 0.09$ & $0.65 \pm 0.07$ & $0.66 \pm 0.07$ \\
 & \textbf{SVC} & 372 & $\bm{1.00 \pm 0.00}$ & $\underline{0.99 \pm 0.00}$ & $\underline{0.99 \pm 0.00}$ & $\underline{0.99 \pm 0.00}$ & $\underline{0.99 \pm 0.00}$ & $\underline{0.99 \pm 0.00}$ \\
 & \textbf{NN} & 460 & $\bm{1.00 \pm 0.00}$ & $\bm{1.00 \pm 0.00}$ & $\bm{1.00 \pm 0.00}$ & $\bm{1.00 \pm 0.00}$ & $\bm{1.00 \pm 0.00}$ & $\bm{1.00 \pm 0.00}$ \\
 & \textbf{NystroemSVM} & 52 & $\bm{1.00 \pm 0.00}$ & $0.98 \pm 0.01$ & $0.98 \pm 0.01$ & $0.98 \pm 0.01$ & $0.98 \pm 0.01$ & $0.98 \pm 0.01$ \\
\midrule
Manifold & McQuack & 2,008 & $0.98 \pm 0.01$ & $0.92 \pm 0.02$ & $0.92 \pm 0.02$ & $0.92 \pm 0.02$ & $0.92 \pm 0.02$ & $0.92 \pm 0.02$ \\
 & VC & 3,024 & $0.93 \pm 0.00$ & $0.86 \pm 0.01$ & $0.86 \pm 0.01$ & $0.86 \pm 0.01$ & $0.86 \pm 0.01$ & $0.86 \pm 0.01$ \\
 & \underline{SVC} & 253 & $\underline{0.99 \pm 0.00}$ & $\underline{0.95 \pm 0.00}$ & $\underline{0.95 \pm 0.00}$ & $\underline{0.95 \pm 0.00}$ & $\underline{0.95 \pm 0.00}$ & $\underline{0.95 \pm 0.00}$ \\
 & \textbf{NN} & 2,016 & $\bm{1.00 \pm 0.00}$ & $\bm{0.96 \pm 0.00}$ & $\bm{0.96 \pm 0.00}$ & $\bm{0.96 \pm 0.00}$ & $\bm{0.96 \pm 0.00}$ & $\bm{0.96 \pm 0.00}$ \\
 & NystroemSVM & 7 & $0.88 \pm 0.03$ & $0.80 \pm 0.03$ & $0.80 \pm 0.03$ & $0.80 \pm 0.03$ & $0.80 \pm 0.03$ & $0.80 \pm 0.03$ \\
\midrule
MNIST & \underline{McQuack} & 6,284 & $\underline{0.92 \pm 0.01}$ & $\underline{0.84 \pm 0.01}$ & $\underline{0.84 \pm 0.01}$ & $\underline{0.84 \pm 0.01}$ & $\underline{0.84 \pm 0.01}$ & $\underline{0.84 \pm 0.01}$ \\
 & VC & 9,432 & $0.89 \pm 0.00$ & $0.80 \pm 0.01$ & $0.80 \pm 0.01$ & $0.80 \pm 0.01$ & $0.80 \pm 0.01$ & $0.80 \pm 0.01$ \\
 & \textbf{SVC} & 527 & $\bm{0.98 \pm 0.00}$ & $\bm{0.91 \pm 0.00}$ & $\bm{0.92 \pm 0.00}$ & $\bm{0.92 \pm 0.00}$ & $\bm{0.92 \pm 0.00}$ & $\bm{0.92 \pm 0.00}$ \\
 & NN & 6,304 & $0.83 \pm 0.19$ & $0.75 \pm 0.27$ & $0.80 \pm 0.17$ & $0.76 \pm 0.23$ & $0.80 \pm 0.15$ & $0.80 \pm 0.17$ \\
 & NystroemSVM & 7 & $0.73 \pm 0.03$ & $0.66 \pm 0.01$ & $0.66 \pm 0.01$ & $0.66 \pm 0.01$ & $0.67 \pm 0.01$ & $0.66 \pm 0.01$ \\
\midrule
MNIST10 & McQuack & 12,556 & $0.96 \pm 0.00$ & $0.77 \pm 0.01$ & $0.76 \pm 0.01$ & $0.76 \pm 0.01$ & $0.76 \pm 0.01$ & $0.76 \pm 0.01$ \\
 & \textbf{VC} & 47,160 & $\bm{0.98 \pm 0.00}$ & $\underline{0.84 \pm 0.00}$ & $\underline{0.83 \pm 0.00}$ & $\underline{0.83 \pm 0.00}$ & $\underline{0.83 \pm 0.00}$ & $\underline{0.83 \pm 0.00}$ \\
 & SVC & 8,964 & $0.96 \pm 0.00$ & $0.75 \pm 0.02$ & $0.70 \pm 0.01$ & $0.69 \pm 0.02$ & $0.69 \pm 0.02$ & $0.70 \pm 0.01$ \\
 & \textbf{NN} & 12,840 & $\bm{0.98 \pm 0.00}$ & $\bm{0.86 \pm 0.01}$ & $\bm{0.86 \pm 0.01}$ & $\bm{0.86 \pm 0.01}$ & $\bm{0.85 \pm 0.01}$ & $\bm{0.86 \pm 0.01}$ \\
 & \underline{NystroemSVM} & 310 & $\underline{0.97 \pm 0.00}$ & $0.81 \pm 0.02$ & $0.81 \pm 0.02$ & $0.80 \pm 0.02$ & $0.80 \pm 0.02$ & $0.81 \pm 0.02$ \\
\midrule
MNIST4 & \textbf{McQuack} & 7,852 & $\bm{1.00 \pm 0.00}$ & $\underline{0.95 \pm 0.02}$ & $\underline{0.95 \pm 0.02}$ & $\underline{0.95 \pm 0.02}$ & $\underline{0.95 \pm 0.02}$ & $\underline{0.95 \pm 0.02}$ \\
 & \textbf{VC} & 18,864 & $\bm{1.00 \pm 0.00}$ & $\underline{0.95 \pm 0.00}$ & $\underline{0.95 \pm 0.00}$ & $\underline{0.95 \pm 0.00}$ & $\underline{0.95 \pm 0.00}$ & $\underline{0.95 \pm 0.00}$ \\
 & SVC & 3,000 & $0.66 \pm 0.08$ & $0.17 \pm 0.14$ & $0.32 \pm 0.10$ & $0.19 \pm 0.12$ & $0.33 \pm 0.09$ & $0.32 \pm 0.10$ \\
 & \textbf{NN} & 7,914 & $\bm{1.00 \pm 0.00}$ & $\bm{0.97 \pm 0.00}$ & $\bm{0.97 \pm 0.00}$ & $\bm{0.97 \pm 0.00}$ & $\bm{0.97 \pm 0.00}$ & $\bm{0.97 \pm 0.00}$ \\
 & \underline{NystroemSVM} & 52 & $\underline{0.98 \pm 0.01}$ & $0.90 \pm 0.03$ & $0.90 \pm 0.03$ & $0.90 \pm 0.03$ & $0.90 \pm 0.03$ & $0.90 \pm 0.03$ \\
\midrule
PM2 & \textbf{McQuack} & 2,060 & $\bm{1.00 \pm 0.00}$ & $\bm{1.00 \pm 0.00}$ & $\bm{1.00 \pm 0.00}$ & $\bm{1.00 \pm 0.00}$ & $\bm{1.00 \pm 0.00}$ & $\bm{1.00 \pm 0.00}$ \\
 & \textbf{VC} & 3,096 & $\bm{1.00 \pm 0.00}$ & $\bm{1.00 \pm 0.00}$ & $\bm{1.00 \pm 0.00}$ & $\bm{1.00 \pm 0.00}$ & $\bm{1.00 \pm 0.00}$ & $\bm{1.00 \pm 0.00}$ \\
 & \textbf{SVC} & 364 & $\bm{1.00 \pm 0.00}$ & $\bm{1.00 \pm 0.00}$ & $\bm{1.00 \pm 0.00}$ & $\bm{1.00 \pm 0.00}$ & $\bm{1.00 \pm 0.00}$ & $\bm{1.00 \pm 0.00}$ \\
 & \textbf{NN} & 2,080 & $\bm{1.00 \pm 0.00}$ & $\bm{1.00 \pm 0.00}$ & $\bm{1.00 \pm 0.00}$ & $\bm{1.00 \pm 0.00}$ & $\bm{1.00 \pm 0.00}$ & $\bm{1.00 \pm 0.00}$ \\
 & \textbf{NystroemSVM} & 7 & $\bm{1.00 \pm 0.01}$ & $\underline{0.99 \pm 0.03}$ & $\underline{0.99 \pm 0.03}$ & $\underline{0.99 \pm 0.03}$ & $\underline{0.99 \pm 0.03}$ & $\underline{0.99 \pm 0.03}$ \\
\midrule
PM4 & \textbf{McQuack} & 2,572 & $\bm{1.00 \pm 0.00}$ & $\bm{1.00 \pm 0.00}$ & $\bm{1.00 \pm 0.00}$ & $\bm{1.00 \pm 0.00}$ & $\bm{1.00 \pm 0.00}$ & $\bm{1.00 \pm 0.00}$ \\
 & \underline{VC} & 6,192 & $\underline{0.99 \pm 0.00}$ & $0.96 \pm 0.01$ & $0.95 \pm 0.01$ & $0.95 \pm 0.01$ & $0.95 \pm 0.01$ & $0.95 \pm 0.01$ \\
 & \textbf{SVC} & 249 & $\bm{1.00 \pm 0.00}$ & $\bm{1.00 \pm 0.00}$ & $\bm{1.00 \pm 0.00}$ & $\bm{1.00 \pm 0.00}$ & $\bm{1.00 \pm 0.00}$ & $\bm{1.00 \pm 0.00}$ \\
 & \textbf{NN} & 2,634 & $\bm{1.00 \pm 0.00}$ & $\bm{1.00 \pm 0.00}$ & $\bm{1.00 \pm 0.00}$ & $\bm{1.00 \pm 0.00}$ & $\bm{1.00 \pm 0.00}$ & $\bm{1.00 \pm 0.00}$ \\
 & \textbf{NystroemSVM} & 52 & $\bm{1.00 \pm 0.00}$ & $\underline{0.97 \pm 0.00}$ & $\underline{0.97 \pm 0.00}$ & $\underline{0.97 \pm 0.00}$ & $\underline{0.97 \pm 0.00}$ & $\underline{0.97 \pm 0.00}$ \\
\midrule
URL & \underline{McQuack} & 662 & $\underline{0.95 \pm 0.00}$ & $\underline{0.89 \pm 0.01}$ & $\underline{0.88 \pm 0.02}$ & $\underline{0.88 \pm 0.02}$ & $\underline{0.89 \pm 0.01}$ & $\underline{0.88 \pm 0.02}$ \\
 & VC & 1,008 & $0.84 \pm 0.01$ & $0.73 \pm 0.01$ & $0.60 \pm 0.04$ & $0.51 \pm 0.08$ & $0.58 \pm 0.04$ & $0.60 \pm 0.04$ \\
 & SVC & 741 & $0.87 \pm 0.00$ & $0.78 \pm 0.00$ & $0.78 \pm 0.00$ & $0.78 \pm 0.00$ & $0.78 \pm 0.00$ & $0.78 \pm 0.00$ \\
 & \textbf{NN} & 664 & $\bm{0.97 \pm 0.00}$ & $\bm{0.93 \pm 0.00}$ & $\bm{0.93 \pm 0.00}$ & $\bm{0.93 \pm 0.00}$ & $\bm{0.93 \pm 0.00}$ & $\bm{0.93 \pm 0.00}$ \\
 & NystroemSVM & 7 & $0.82 \pm 0.03$ & $0.72 \pm 0.04$ & $0.72 \pm 0.04$ & $0.71 \pm 0.04$ & $0.72 \pm 0.04$ & $0.72 \pm 0.04$ \\
\bottomrule
\end{tabular}
\end{table*}
\endgroup

\clearpage

\subsection{Hyperparameters}  \label{app:hyperparameters}

\subsubsection{McQuack} \label{subsec:appC:McQuack_hyperparameters}

\begin{table}[H]
\centering
\caption{Hyperparameter search space of McQuack for the QML Benchmark suite.}
\label{app:tab:qml_benchmarks_hyperparams}
\begin{tabular}{ll}
\toprule
\textbf{Parameter} & \textbf{Values} \\
\midrule
n\_qubits & 6 \\
n\_repetitions & 1, 5 \\
epochs & 500 \\
epochs\_ka & 1, 2 \\
epochs\_co & 1, 2 \\
lr\_ka & 1e-3, 1e-2, 1e-1 \\
lr\_co & 1e-3, 1e-2, 1e-1 \\
decay\_rate & 1.0 \\
reg\_param\_weights & 0.0, 1e-3, 1e-2 \\
reg\_param\_bias & 0.0, 1e-3 \\
reg\_param\_co & 1e-3 \\
batch\_size & 32 \\
patience & 50 \\
init\_strategy & wx+b=0 \\
init\_small & True \\
\bottomrule
\end{tabular}
\end{table}

\begin{table}[H]
\centering
\caption{Hyperparameter search space of McQuack for the real-world datasets.}
\label{app:tab:real_world_hyperparams}
\begin{tabular}{ll}
\toprule
\textbf{Parameter} & \textbf{Values} \\
\midrule
n\_qubits & 6 \\
n\_repetitions & 1 \\ 
epochs & 200 \\
epochs\_ka & 1 \\
epochs\_co & 1 \\
lr\_ka & 1e-3 \\
lr\_co & 1e-3 \\
decay\_rate & 0.99 \\
reg\_param\_weights & 1e-3, 1e-2 \\
reg\_param\_bias & 1e-3, 1e-2 \\ 
reg\_param\_co & 1e-3 \\
batch\_size & 100 \\
patience & 5 \\
init\_strategy & default \\
init\_small & True \\
\bottomrule
\end{tabular}
\end{table}

\subsubsection{Variational Circuits}
An ensemble of \acp{VC} was used as quantum baseline for the real-world datasets. This model measures the first two qubits. Each of the expectation values is the unnormalized logit for a class (or one class vs. all other classes in the multiclass case). The Cross Entropy is used as loss function. 
The formal definition of the ensemble is given in \cref{eq:ensemble_vc_formal} in \cref{subsec:map_to_vc} and a single classifier is given by
\begin{align*}
h_m(\boldsymbol{x_i}; \boldsymbol{w_m}, \boldsymbol{b_m}) = 
\left(\begin{array}{cc}
\langle 0 | U(\boldsymbol{x_i}; \boldsymbol{w_m}, \boldsymbol{b_m})^{\dagger} Z_{0} U(\boldsymbol{x_i}; \boldsymbol{w_m}, \boldsymbol{b_m}) | 0 \rangle \\
\langle 0 | U(\boldsymbol{x_i}; \boldsymbol{w_m}, \boldsymbol{b_m})^{\dagger} Z_{1} U(\boldsymbol{x_i}; \boldsymbol{w_m}, \boldsymbol{b_m}) | 0 \rangle
\end{array}\right),
\end{align*}
with unitary $U$ as defined in \cref{eq:full_U_definition} in \cref{Circuit Design and Data Encoding}.
Each classifier is trained individually and after training, only the PauliZ-0 expectation value is used in \cref{eq:ensemble_vc_formal} in \cref{subsec:map_to_vc}.
The prediction of $H$ is analogous as defined in \cref{subsec:map_to_vc}

The model uses the same hyperparameters as McQuack for the real-world datasets and the learning rate was optimized in a grid search with search space $\{10^{-2}, \, 10^{-3}\}$.

\subsubsection{SVC}
A vanilla Support Vector Classifier was used as kernel benchmark. Multiclass classification is performed according to the one-vs-one scheme\footnote{sklearn's implementation uses by default \texttt{decision\_function\_shape = ovr}, however \texttt{ovo} is always used as a multi-class strategy to train models. See \url{https://scikit-learn.org/1.4/modules/generated/sklearn.svm.SVC.html}}. 
The hyperparameters were optimized in a grid search from the parameter space 
\begin{verbatim}
    {'svc_C': {0.1, 1, 10, 100}, 'svc_gamma': {0.001, 0.01, 0.1, 1}}
\end{verbatim}
and all other hyperparameters were the default ones from sklearn.
The number of parameters reported in \cref{tab:results_real_world_ds} is the number of dual coefficients of the model.

\subsubsection{NystroemSVM}
\textsc{NystroemSVM} is a classical kernel method with linear scaling in training set size. This is achieved by approximating an RBF kernel through the Nystroem method.
We first intended to have the same number of kernel entries in \textsc{NystroemSVM} and McQuack by choosing the same number of landmark points in \textsc{NystroemSVM} as McQuack has centroids, i.e. $M$ landmark points. However, due to the random sampling it is not guaranteed that the class distribution of the landmark points represents the class distribution in the training set.
Hence, we increased the number of landmark points in \textsc{NystroemSVM} threefold to $3M$ so that the probability of sampling at least one landmark point of each class in a balanced binary dataset is approximately $97\%$. 
Multiclass classification is performed according to the one-vs-rest scheme. 
The hyperparameter \texttt{scale} was optimized in a grid search with search space 
\begin{verbatim}
    {'scale' : {0.1, 0.2, 0.5, 1.0, 2.0, 5.0, 10.0, 20.0, 50.0, 100.0}}
\end{verbatim}
for the real-world datasets and 
\begin{verbatim}
    {'scale' : {0.1, 0.2, 0.5, 1.0, 2.0, 5.0, 10.0}}
\end{verbatim}
for the QML benchmark suite.

\subsubsection{Neural Net}
The final baseline is a two-layer feedforward neural network. 
The number of neurons per layer was selected such that the total number of trainable parameters is as close as possible but not smaller than the number of parameters of McQuack.
The activation function has been optimized for each dataset in a grid search with search space \{ \texttt{relu}, \texttt{tanh}, \texttt{sigmoid} \}. The models are trained for 400 epochs with a learning rate of $0.001$.
\clearpage
\section{Quantum Kernels} \label{sec:appD:quantum_kernels}
In the following, the preliminaries of \ac{QKM} required to understand McQuack are introduced.

\ac{QKM}s are of particular interest in the \ac{QML} community because the mathematical framework of \ac{QC} is similar to the one of kernel methods. Both analyse data in high-dimensional Hilbert spaces which can be accessed only through inner products revealed by measurements \cite{schuld2021supervisedquantummachinelearning}. 

In general, \ac{QML} models use a non-linear feature map $\Phi : \mathcal{X} \to \mathcal{F}$ to encode data $\boldsymbol{x} \in \mathcal{X}$ into vectors of a potentially inaccessibly large Hilbert space, called feature space $\mathcal{F}$, without the need of ever computing an explicit numerical representation of these vectors. 
Once the data has been encoded in the Hilbert (or feature) space, the model becomes linear in this space, and hence the data-encoding routine is particularly important \cite{Quantum_Computing_Schuld2021, Havl_ek_2019}.

The feature map $\Phi$ is realized through unitary operations $U(\boldsymbol{x}_i)$ composed of quantum gates acting on $n$-qubits initialized in the zero-state. The quantum state $\ket{\psi_i} = \ket{\psi(\boldsymbol{x}_i)}$ of the system after encoding a data vector $\boldsymbol{x}_i$ is

\begin{align*}
\Phi (\boldsymbol{x}_i)  = U(\boldsymbol{x}_i) \ket{0}^{\otimes n} = \ket{\psi_i}.
\end{align*}

Kernels are known from classical machine learning, where they are real- or complex-valued positive definite functions of two data points, i.e. $\kappa : \mathcal{X} \times \mathcal{X} \rightarrow \mathbb{K}$, where $\mathbb{K} \in \{\mathbb{R}, \mathbb{C}\}$.
This definition can be extended to the quantum case, where the fidelity kernel $k$ between two data-encoding pure quantum states $\psi_{i}$ and $\psi_{j}$ is calculated from the fidelity $F$ between these states
\begin{align}
     k(\boldsymbol{x}_i, \boldsymbol{x}_j) &= F(\psi_{i}, \psi_{j}) = \vert \braket{\psi_{j}|\psi_{i}} \vert ^2 \label{eq:quantum_kernel_sim} \\
     &= \vert \bra{0^{\otimes n}} U^\dagger(\boldsymbol{x}_j) U(\boldsymbol{x}_i) \ket{0^{\otimes n}} \vert ^2 \label{eq:quantum_kernel_hardware}
\end{align}
with data encoding unitaries $U(\boldsymbol{x}_i)$ and $U^\dag(\boldsymbol{x}_j)$, where $U^\dag$ denotes the conjugate transpose of $U$, and $\braket{\cdot|\cdot}$ is the inner product. 
This quantum kernel serves as a similarity measure between the states of two encoded samples:
If both samples are identical, i.e. $\boldsymbol{x}_i = \boldsymbol{x}_j$, equivalent to $\psi_{i} = \psi_{j}$, the kernel equation \eqref{eq:quantum_kernel_sim} simplifies to
\begin{align*}
     k(\boldsymbol{x}_i, \boldsymbol{x}_j) = k(\boldsymbol{x}_i, \boldsymbol{x}_i) = F(\psi_{i}, \psi_{i}) = \vert \braket{\psi_{i}|\psi_{i}} \vert ^2  = 1.
\end{align*}

On the other hand, if the encoded states $\psi_{i}$ and $\psi_{j}$ are orthogonal, the kernel will evaluate to 
\begin{align*}
     k(\boldsymbol{x}_i, \boldsymbol{x}_j) &= F(\psi_{i}, \psi_{j}) = \vert \braket{\psi_{j}|\psi_{i}} \vert ^2 = 0.
\end{align*}

\begin{figure}[htb]
\centerline{\includegraphics[width=.35\textwidth]{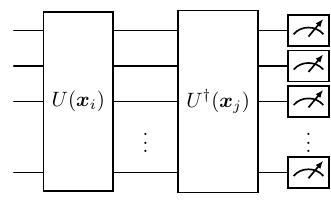}}
\caption{Architecture of the circuit if executed on hardware. The kernel entry $K_{ij} = k (\boldsymbol{x}_i, \boldsymbol{x}_j)$ for samples $i$ and $j$ is the probability of measuring the all-zero bit string.}
\label{fig:circuit}
\end{figure}

On a quantum computer, a quantum kernel can be implemented as an $n$-qubit circuit that consists of a unitary $U(\boldsymbol{x}_i)$, encoding a single sample, followed by the unitary's complex conjugate $U^\dagger(\boldsymbol{x}_j)$ encoding another sample, and a measurement of all qubits, as shown in Fig.~\ref{fig:circuit}. The kernel value of the two samples is then obtained as the probability of measuring the all-zero state as given in equation \eqref{eq:quantum_kernel_hardware}. If a state vector simulator is used, the kernel value of two samples $\boldsymbol{x}_i$ and $\boldsymbol{x}_j$ is the fidelity of the states after application of the unitary $U(\boldsymbol{x}_i)$, respectively $U(\boldsymbol{x}_j)$, as given in \eqref{eq:quantum_kernel_sim}.

\end{document}